\documentclass[aip,jcp,amsmath,amssymb,reprint,floatfix]{revtex4-1}
\usepackage[dvipdfm]{graphicx}
\usepackage{dcolumn}
\usepackage{bm}
\usepackage{amsmath,amsthm,amssymb}
\usepackage{cases,mathrsfs}
\usepackage{ulem}
\usepackage{wrapfig}
\usepackage{ascmac,paralist}
\usepackage{slashbox}
\usepackage{latexsym}
\usepackage{makeidx}
\usepackage{braket}
\def\yfigcaption#1{
\def\@currentlabel{ } 
 #1
\addcontentsline{lof}{figure}{\protect\numberline {\@currentlabel}{\ignorespaces #1}}
}
\newcommand{\iGa}{{\it \Gamma}}
\newcommand{\dis}{\displaystyle}
\begin{document}
\title{Relaxation Mode Analysis and Markov State Relaxation Mode Analysis for Chignolin in Aqueous Solution near a Transition Temperature}
\author{Ayori Mitsutake}
\email{ayori@rk.phys.keio.ac.jp}
\affiliation{
Department of Physics, Faculty of Science and Technology, Keio University,\\
Yokohama, Kanagawa 223-8522, Japan}
\affiliation{
JST, PRESTO, Yokohama, Kanagawa 223-8522, Japan}
\author{Hiroshi Takano}
\email{takano@rk.phys.keio.ac.jp}
\affiliation{
Department of Physics, Faculty of Science and Technology, Keio University,\\
Yokohama, Kanagawa 223-8522, Japan}
\begin{abstract}
It is important to extract reaction coordinates or order parameters
from protein simulations
in order to
investigate the local minimum-energy states and the transitions between them.
The most popular method to obtain such data is principal component analysis,
which extracts modes of large conformational fluctuations
around an average structure.
We recently applied relaxation mode analysis for protein systems, which
approximately estimates the slow relaxation modes and times from a simulation and enables investigations of the dynamic properties underlying the structural fluctuations of proteins.
In this study, we apply this relaxation mode analysis
to extract reaction coordinates for a system
in which there are large conformational changes
such as those commonly observed in protein folding/unfolding.
We performed a 750-ns simulation of chignolin protein
near its folding transition temperature, and observed many
transitions between the most stable, misfolded, intermediate, and unfolded states.
We then applied principal component analysis
and relaxation mode analysis to the system.
In the relaxation mode analysis,
we could automatically extract good reaction coordinates.
The free-energy surfaces provide a clearer understanding of
the transitions not only between local minimum-energy states
but also between the folded and unfolded states,
even though the simulation involved large conformational changes.
Moreover, we propose a new analysis method called Markov state relaxation mode analysis.
We applied the new method to states with slow relaxation, which are defined by the free-energy surface obtained in the relaxation mode analysis.
Finally, the relaxation times of the states obtained with a simple Markov state model and the proposed Markov state relaxation mode analysis are compared and discussed.
\end{abstract}
\maketitle
\section{INTRODUCTION}
\par
Biopolymers have flexible structures and show variable functions.
These functions are derived not only from the structure but also from the dynamics of the structural fluctuations themselves.
Therefore,
identifying the dynamic properties of the structural fluctuations of biopolymers is
important for understanding the interrelationship between their movements and functions.
Classical molecular dynamics (MD) simulation is a popular and powerful method for describing the
structure, dynamics, and function of biomolecules in microscopic detail.
Recent technological advances have allowed for simulations to be carried out on the timescales of the order of milliseconds (see reviews by Refs.\ \citenum{Shaw,Pande}).
As longer and larger MD simulations are now being performed, it has become more important to develop methods for extracting
the most ``essential'' modes from the trajectory \cite{nphys}.
Indeed, the development of a method to reduce the huge number of degrees of freedom of coordinates to a few collective degrees of freedom is
an active field of theoretical research.
\par
Principal component analysis (PCA), also called quasi-harmonic analysis or the essential dynamics method \cite{KHG,IK,AA,G,ALB,HKHG,qh},
is one of the most popular methods for analyzing the static properties of the fluctuations of structures.
A system's fluctuations can be described in terms of only a few principal components (PCs).
Moreover, this method has been widely used to extract the few important collective modes of a biomolecule,
which may serve as relevant coordinates of the free-energy surface.
However, if the simulation involves large conformational changes such as folding/unfolding simulations,
and the conformational changes between local minimum-energy states are
small compared with those between the folded and unfolded states,
it is difficult for PCA to extract the effective modes or order parameters
to identify the local minimum-energy states.
The root mean square distance (RMSD) from a reference structure,
the radius of gyration, or selected distances, among others, are often used as order
parameters to construct free-energy surfaces and analyze the structures obtained from a simulation.
The selection of these order parameters depends on the simulation system.
To identify the local minimum-energy states from the simulation,
suitable order parameters must be considered,
which depend on the simulation system.
Therefore, given the recent possibility for performing longer MD simulations, a new analysis method must be able to extract suitable order parameters for identifying local minimum-energy states automatically, and for investigating the dynamics and kinetics of proteins.
\par
Relaxation mode analysis (RMA) was developed to investigate the ``dynamic'' properties of spin systems\cite{TM} and homopolymer systems \cite{KHT,HKT}, and has been applied to various polymer systems \cite{HT1,HaKT,HIT,HT2,HT3,ST,IT,IT2,IHT} to investigate their slow relaxation dynamics \cite{dG,DE}.
Recently, RMA has also been applied to biomolecular systems \cite{MIT,NMT}.
The relaxation modes and rates are given as left eigenfunctions and eigenvalues of
the time-evolution
operator of the master equation of the system, respectively \cite{TM,KHT,HKT}.
The equilibrium time correlation functions of the relaxation modes, $\{X_p\}$, satisfy
\begin{equation}
\langle X_p(t)X_q(0) \rangle=\delta_{p,q} e^{-\lambda_p t},
\end{equation}
where $\langle X_p(t)X_q(0) \rangle$ denotes the equilibrium correlation of $X_p$ at time $t$ and
$X_q$ at time $0$, and $\lambda_p$ represents the relaxation rate of $X_p$.
\par
In RMA, we calculate the approximate relaxation modes and rates from
a simulation.
RMA is formulated as a variational problem equivalent to the eigenvalue problem of the time-evolution operator of the system,
where a relaxation mode $X_p$ is approximated by a trial function, which
is constructed as a linear combination of relevant physical quantities that are time-evolved for $t_0/2$.
The parameter $t_0$ is introduced in order to reduce the relative weight of the faster modes contained in the
physical quantities.
The optimization of the normalized equilibrium autocorrelation function of
the trial function time-displaced by $\tau$ leads to
the generalized eigenvalue problem.
In practice, the time correlation matrices of structural fluctuations for two different times ($t_0$ and $t_0+\tau$) are calculated through simulations.
Then, by solving a generalized eigenvalue problem for these matrices, the relaxation rates and modes are obtained from the eigenvalues and eigenvectors, respectively.
Thus, we call this method the RMA method using a single evolution time: $t_0/2$.
\par
Recently, RMA was applied to a Monte Carlo simulation
for a heteropolymer peptide system with a small number of degrees of freedom\cite{MIT}, and its effectiveness was demonstrated.
In the applications of RMA to homopolymer systems \cite{KHT,HKT,HT1,HaKT,HIT,HT2,HT3,ST}, the translational degrees of freedom are removed from the conformations of the polymer sampled in a simulation.
However, the rotational degrees of freedom are retained because
the slow rotational relaxation of the polymer is important in polymer physics.
In contrast, for heteropolymer biomolecule systems,
both the translational and rotational degrees of freedom
are removed from the sampled conformations of a biomolecule to allow for investigation of the structural fluctuations around
its average conformation.
In our previous report\cite{MIT},
we explained how to treat such sampled conformations
with RMA, and
succeeded in applying RMA to a heteropolymer system.
Moreover, we demonstrated the effectiveness of RMA
to investigate the transitions
between some local minimum-energy states
by calculating the free-energy surfaces for
relaxation modes.
\par
Although RMA is a powerful method for extracting
slow modes, it requires a relatively high level of statistical precision of the
time correlation matrices.
Owing to this requirement, RMA requires a long simulation where many transitions between local minimum-energy states occur,
which may not be a problem because such very long simulations have become feasible in recent years, as mentioned above.
Moreover, RMA cannot handle a large number of degrees of freedom directly.
Therefore, reduction of the degrees of freedom is necessary
for application of RMA to protein systems.
To achieve this,
we proposed a new analysis method, which is referred to as
the principal component relaxation mode analysis (PCRMA) method.
In this method, PCA is carried out first and then RMA is applied to
a small number of principal components with large fluctuations.
This method can systematically reduce the degrees of freedom.
We also proposed the RMA method using multiple evolution times,
which can reduce the contribution of the fast modes efficiently.
In our previous report \cite{NMT},
we explained
the combination of two proposed methods involving the
PCRMA method using multiple evolution times,
and demonstrated its
applicability to an all-atom MD simulation of human lysozyme in aqueous solution.
\par
Recently, methods to analyze the dynamics and kinetics of protein simulations have been developed.
In particular, the Markov state model has been developed (see Refs.\ \citenum{SPS,SP,CSPD,CDSPSW,NHSS,BH} and reviews \cite{MSM,NoeRev,PandeRev} and references cited therein) and used for many protein systems.
The Markov state model can analyze transitions between
local minimum-energy states, which are identified from clustering analysis methods.
This is a powerful method for analyzing dynamics in the context of both long and short simulations of proteins.
As mentioned above, in RMA, the relaxation modes and rates are given as left eigenfunctions
and eigenvalues of the time-evolution operator of the master equation of the system, respectively \cite{TM,KHT,HKT}
(see Appendix A).
From this point of view, RMA is related to Markov state models
(see \ref{MSRMA}, Appendices A and B, and Ref.\ \citenum{MSM2}).
\par
An analysis method of protein dynamics based on MD simulation was proposed \cite{Fuchigami} that separates linearly superimposed statistically
independent signals by using time correlation functions,
and was thus designated as time-structure based independent component analysis
(tICA).
The method is closely related to RMA,
in that it determines statistically independent modes by
solving the generalized eigenvalue problem with $t_0 = 0$.
Recently, the combination method of tICA and a Markov state model was also proposed \cite{MSM2,MSMPANDE}.
In Ref.\ \citenum{MSM2}, a Markov state model was constructed from clustering in the subspace determined by tICA, which was calculated using a method similar to PCRMA.
\par
In this study, we applied PCA and RMA to analyze the folding trajectories of a peptide, chignolin, near its folding transition temperature.
As mentioned above, PCA cannot extract order parameters to easily identify the local minimum-energy states for a system with large conformational changes.
This is because PCA extracts the modes with large structural fluctuations around an average structure, and the
PC modes with large structural fluctuations do not correspond to the transitions between the local minimum-energy states.
On the other hand, RMA can extract slow relaxation modes.
The local minimum-energy states are usually stable and
the system thus remains in this state for a long time during a simulation.
The order parameters with slow relaxation may correspond to the directions between local minimum-energy states.
Thus, slow relaxation modes may be suitable order parameters
to identify local minimum-energy states and the transitions between them.
\par
To test this model, we performed a simulation of chignolin near its transition temperature.
Chignolin consists of a 10 amino-acid polypeptide that adopts a $\beta$-hairpin turn structure, and is the smallest artificial protein \cite{chig}, which has been used for testing new simulation algorithms.
Indeed, there are many simulations for chignolin reported to date\cite{terada,taiji,haradakitao,Best,Oku}.
Previous research has shown that
chignolin has a stable state (native structure) and a misfolded state.
The native and misfolded structures are hairpin-like structures (see Fig.~\ref{Fig9}). They have a common turn structure from Asp3 to Glu5 and have slightly
different hydrogen bond patterns.
Furthermore, the stability of the two states, i.e., how often these two states are observed during a simulation, has been found to depend on the force field \cite{Best}.
We considered that by increasing temperature, many transitions between the two states may occur within a several hundred-nanosecond simulation.
Thus, we performed the simulation near the protein's transition temperature,
which corresponds to a long simulation at room temperature.
For estimation of the transition temperature,
we refer to the results at a pressure of 1 atm obtained by the simulations in 
Ref.\ \citenum{Oku}.
We calculated the free-energy surfaces of the coordinates of PCA and RMA and
show the effectiveness of RMA for extracting order parameters of the protein system with large conformational changes.
\par
We also explain the relationship between
RMA and Markov state models and propose a new analysis method called Markov state relaxation mode analysis (MSRMA).
We evaluate the relaxation times of the states, which are defined by the free-energy surface from RMA, using a simple Markov state model and the new MSRMA method.
\section{METHODS}
\subsection{Principal component analysis}
\label{SubSecPCA}
PCA is a well-known method for analyzing the static properties of protein structural fluctuations obtained in a protein simulation system  \cite{KHG,IK,AA,G,HKHG,ALB,qh}.

\par
We now consider a biopolymer composed of $N$ atoms.
We assume that ${\bm R}$ is a $3N$-dimensional column vector that consists of a set of coordinates of atoms
relative to their average coordinates.
Namely,
\begin{equation}
{\bm R}^{\rm T}=( {{\bm r}_1^{\prime}}^ {\rm T},{{\bm r}_2^{\prime}}^{\rm T},\ldots,{{\bm r}_N^{\prime}}^{\rm T} )
=(x_1^{\prime},y_1^{\prime},z_1^{\prime},\ldots,x_N^{\prime},y_N^{\prime},z_N^{\prime})
\label{DefOfR}
\end{equation}
with
\begin{equation}
{\bm r}_i^{\prime}={\bm r}_i - \langle{\bm r}_i \rangle~,
\label{eqP1}
\end{equation}
where ${\bm r}_i$ is the coordinate of the $i$th atom of the biopolymer
and $\langle {\bm r}_i \rangle$ is its average coordinate.
\par
In PCA, the eigenvalue problem is solved as
\begin{equation}
C {\bm F}_n = \Lambda_n {\bm F}_n~~~~\mbox{with}~~~~
{\bm F}_m^{\rm T}{\bm F}_n = \delta_{m,n}.
\label{pri1}
\end{equation}
Here, $C = \langle {\bm R}{\bm R}^{\rm T}\rangle$ is
the $3N \times 3N$ variance-covariance matrix:
$ C = \left( C_{i,j} \right)$ with
$ C_{i,j} = \left\langle R_i R_j \right\rangle$.
We now set the indices of the eigenvalues so that
the relation $\Lambda_1 \ge \Lambda_2 \ge \cdots \ge \Lambda_{3N}$ holds.
The eigenvector ${\bm F_n}$ with the eigenvalue $\Lambda_n$ is referred to
as the $n$th principal component axis.
Note that $\Lambda_{3N-5}=\Lambda_{3N-4}=\cdots=\Lambda_{3N}=0$
because the translational and rotational degrees of freedom are removed.
The coordinate ${\bm R}$ can be expanded in terms of
the PCA eigenvectors:
\begin{align}
{\bm R} = \sum_{n=1}^{3N-6} \Phi_{n} {\bm F}_{n}
~~~{\rm with}~~~
\Phi_{n} = {\bm F}_n^{\rm T} {\bm R}.
\label{pri3}
\end{align}
Here, $\Phi_{n}$ is called the $n$th principal component.
The variance of $\Phi_{n}$ is given by $\Lambda_n$.
\subsection{Relaxation mode analysis}
\label{RMA}
The RMA method was developed
to identify slow motions in random spin systems and homopolymer systems.
In RMA, we consider the following physical quantities:
\begin{equation}
\langle \phi_n(t)\phi_m(0) \rangle=\delta_{n,m}{\rm e}^{-\lambda_n t}~,
\end{equation}
where $\phi_n$ and $\lambda_n$ are the
relaxation mode and its relaxation rate, respectively.
Here, $\langle A(t)B(0) \rangle$ denotes the equilibrium time correlation
function of physical quantity $A$ at time $t$ and $B$ at time $0$.
\par
In the following, we briefly explain how to use RMA to extract slow relaxation modes and their relaxation rates from
an MD simulation that satisfies the detailed balance condition.
Here, we are only dealing with position coordinates because
the relaxation of velocities (on a picosecond time scale) is usually faster than that of slow collective modes for coordinates (on a nanosecond time scale).
In this case, the calculation process is the same as that for Monte Carlo simulations
(see Ref.\ \citenum{MIT}).
Note that we describe the theoretical background of RMA
in Appendix A and the application of the RMA method for an MD simulation
in detail in Appendix B.
\par
We use the following function as an approximate relaxation mode:
\begin{equation}
X_{p}=
{{\bm f}_{p}}^{\rm T} {\rm e}^{-\iGa^\dagger t_0/2}{\bm R}
\label{eB5}
\end{equation}
Here, ${\bm R}$ is given by Eq.~(\ref{pri3})
and
$\iGa$ is the time-evolution operator of the master equation
of the system (see Appendices A and B).
The operator ${\rm e}^{-\iGa^{\dagger} t_0/2}$
is used to reduce the contributions of faster modes in ${\bm R}$, and the
trial function becomes a better approximation as $t_0$ becomes larger \cite{MIT}.
Note that Eq.\ (\ref{eB5}) is the same as Eq.\ (30) of Ref.\ \citenum{MIT}.
$X_p$ is a trial function, which is constructed as a linear combination of the expectation value of ${\bm R}$ after a period $t_0/2$.
\par
For the trial functions, the variational problem is
equivalent to a generalized eigenvalue problem
\begin{equation}
C(t_0+\tau){\bm f}_p
= {\rm e}^{-\lambda_p \tau}
 C(t_0){\bm f}_p~,
\end{equation}
where $\lambda_p$ is the relaxation rate corresponding to $X_p$.
The matrix $C(t)$ is defined by
\begin{equation}
C(t)=
\langle {\bm R}(t) {\bm R}^{\rm T}(0) \rangle.
\label{eB10}
\end{equation}
The orthonormal condition for $X_{p}$
is written as
\begin{equation}
{\bm f}_p^{\rm T}
C(t_0)
{{\bm f}_q}
=\delta_{p,q}.
\label{eB11}
\end{equation}
Note that
the number of meaningful relaxation modes is $3N-6$
because the translational and rotational degrees of freedom are removed \cite{MIT}.
\par
The time correlation functions of the approximate relaxation modes obtained
can thus be written as
\begin{equation}
\langle X_p(t)X_q(0) \rangle \simeq \delta_{p,q}{\rm e}^{-\lambda_p t}~.
\label{B11b}
\end{equation}
This relation holds exactly for
$t=0$ and $t=\tau$ and is expected to hold approximately
for other values of $t \ge 0$.
\par
From the orthonormal condition, the inverse transformation is derived as
\begin{equation}
{\rm e}^{-\Gamma^\dagger t_0/2}{\bm R}
=
\sum_{p=1}^{3N-6} {\bm g}_p X_p,
\label{B12}
\end{equation}
where
\begin{equation}
{\bm g}_p = C(t_0) {\bm f}_p.
\label{B12b}
\end{equation}
\par
The time correlation functions of ${\bm R}$ are given as
\begin{align}
\displaystyle C(t)&=\langle {\bm R}(t) {\bm R}^{\rm T}(0) \rangle
\nonumber\\
&=  \displaystyle \sum_{p=1}^{3N-6} \sum_{q=1}^{3N-6} {\bm g}_p {\bm g}^{\rm T}_q \left<X_p(t-t_0)X_q(0) \right>
\nonumber\\
& \simeq  \displaystyle \sum_{p=1}^{3N-6} {\bm g}_{p} {\bm g}^{\rm T}_{p} {\rm e}^{-\lambda_p (t-t_0)}
\nonumber\\
&= \displaystyle \sum_{p=1}^{3N-6} \tilde{{\bm g}}_{p} \tilde{{\bm g}}^{\rm T}_{p} {\rm e}^{-\lambda_p t}
\label{B13}
\end{align}
for $t \ge t_0$.
Here,
\begin{equation}
\tilde{{\bm g}}_{p}=e^{\lambda_p t_0/2} {\bm g}_{p}~.
\end{equation}
Equations (\ref{B11b}) and (\ref{B13}) lead to the relaxation mode
expansion of ${\bm R}$:
\begin{equation}
{\bm R} \simeq  \sum_{p=1}^{3N-6} {\tilde{\bm g}}_{p}X_p ~.
\label{B15}
\end{equation}
Because we are dealing with position coordinates only,
$C(t)$ is a symmetric matrix owing to
the detailed balance condition \cite{R},
and
if $t_0$ is sufficiently large,
the relaxation rates $\lambda_p$ are real and positive,
which correspond to pure relaxation.
Then, we set the indices of $\lambda_p$
so that $0 < \lambda_1 \le \lambda_2 \le \cdots$ holds.
On the basis of the
approximate relaxation modes and rates, the correlation matrix $C(t)$ with $t \ge t_0$ can be calculated.
The accuracy of the present estimation of the relaxation modes and rates can be examined through comparison of the correlation matrices $C(t)$ calculated by the present method with those estimated directly by other means. 
In general, the time-displaced autocorrelations $C_{i,i}(t)$ are compared to check the obtained relaxation modes and rates.
\subsection{Calculation of the free-energy surface}
In PCA, from the probability density $P(\Phi_{p},\Phi_{q})$
of $\Phi_{p}$ and $\Phi_{q}$,
the dimensionless free energy,
which is the free energy divided by $k_{\rm B}T$,
along the $p$th and $q$th principal component axes is calculated as
\begin{equation}
F(\Phi_{p},\Phi_{q}) = - \ln P(\Phi_{p},\Phi_{q}).
\label{FreeEnergyPCA}
\end{equation}
Here, $k_{\rm B}$ and $T$ denote the Boltzmann constant
and the temperature of the system, respectively.
In the following, we abbreviate ``dimensionless free energy''
as ``free energy'' for simplicity.
The indices
$p$ and $q$ are usually set to numbers corresponding to large eigenvalues (e.g., $p=1$ and $q=2$).
\par
In RMA, the quantity $Y_p$,
corresponding to $\Phi_p$ in PCA,
is defined by
\begin{equation}
Y_p = X_p |{\tilde {\bm g}}_p|~.
\label{DefOfYp}
\end{equation}
Then,
the free-energy surface as a function of $Y_p$ and  $Y_q$ is calculated as
\begin{equation}
F(Y_p,Y_q)
= -\ln P(Y_p,Y_q),
\label{freerma}
\end{equation}
where
$ P(Y_p,Y_q) $
denotes the probability density of $Y_p$ and $Y_q$.
Here, $X_p$ is calculated from ${\bm R}$ as follows.
Because of Eqs.\ (\ref{eB11}) and (\ref{B12b}),
${\bm f}_p^{\rm T}{\bm g}_q=\delta_{p,q}$ holds,
which leads to
$\displaystyle
{\bm f}_p^{\rm T}\tilde{\bm g}_q={\rm e}^{\lambda_p t_0 /2}\delta_{p,q}$.
Therefore,
by multiplying ${\bm f}_p^{\rm T}$ to both sides of Eq.\ (\ref{B15}),
$X_p$ is given as a function of ${\bm R}$ as
\begin{equation}
	X_p = \tilde{\bm f}_p^{\rm T}{\bm R}
\end{equation}
with
\begin{equation}
	\tilde{\bm f}_p ={\rm e}^{-\lambda_p t_0 /2}{\bm f}_p.
\end{equation}
\par
\subsection{Markov State Relaxation Mode Analysis}
\label{MSRMA}
In this subsection, we consider the relation between
RMA and Markov state models, and propose the new method of MSRMA.
In the simplest Markov state model,
the phase space of the system,
where only the position coordinates are considered,
is divided
into clusters (subsets) $S_i$, $i=1, \ldots, n$.
First, the joint probability
$\bar{P}_{i,j}(\tau)=P( Q \in S_i, \tau; Q \in S_j , 0)$
that the state of the system $Q$ is in the $j$th cluster at time $0$
and is in the $i$th cluster at time $\tau >0$
is calculated in a simulation.
Second, the transition probability
$\bar{T}_{i,j}(\tau)$ that
the state of the system is found in the $i$th cluster after time $\tau$
starting from a state in the $j$th cluster
is calculated by
\begin{equation}
\bar{T}_{i,j}(\tau) = \bar{P}_{i,j}(\tau)/\bar{p}_j,
\label{T=P/p}
\end{equation}
where $\bar{p}_j = P(Q \in S_j)$ is the probability
that the state of the system is found in the $j$th cluster,
which is estimated in the simulation.
Then, by solving the eigenvalue problem
\begin{equation}
	{\bar{\bm f}_p}{}^{\rm T} \bar{T}(\tau) = {\bar{\bm f}_p}{}^{\rm T} \bar{\Lambda}_p
\label{fTT}
\end{equation}
for the transition matrix
$\bar{T}(\tau) = (\bar{T}_{i,j}(\tau))$,
the $p$th eigenvector $\bar{\bm f}_p$ and its eigenvalue $\bar{\Lambda}_p$
are obtained.
The eigenvector $\bar{\bm f}_1 \propto (1,1,\ldots,1)^{\rm T}$ corresponds to
the equilibrium state and its eigenvalue $\bar{\Lambda}_1 = 1$.
Other eigenvectors $\bar{\bm f}_p$ represent structural transitions and
the corresponding eigenvalues $\bar{\Lambda}_p$ give
their relaxational timescales $\bar{\tau}_p$ as
\begin{equation}
	\bar{\tau}_p = - \frac{\tau}{\ln \bar{\Lambda}_p}.
\label{MSRMAre}
\end{equation}
Note that in the Markov description, it is important that the states are defined in a kinetically meaningful way\cite{SPS,MSM2}.
We need to define the states that are classified by order parameters representing the dynamics and kinetics of the system.
Even with a good choice of states, in order for a Markov description of the process to be accurate,
the time interval $\tau$ should also be chosen carefully.
In other words, for a Markov description to work, the time interval of the transition matrix $\tau$ must be chosen appropriately so that it is
as large as the slowest relaxation time of the states.
When plotting ${\bar \tau}_p$ as a function of $\tau$, ${\bar \tau}_p$ slowly converges to the appropriate time scale
when $\tau$ is increased.
In addition, when a much longer $\tau$ than the
slowest relaxation time of the states is used, the Markov state model is not expected to be accurate.
Thus, we usually set the time interval $\tau$ to the value when the variation of $\tau_p$ is sufficiently flat \cite{SPS,MSM2}.
\par
The above-mentioned procedure of the Markov state model
is related to the following procedure of RMA.
We call the following procedure the MSRMA.
Similar to the description above in Section \ref{RMA},
we consider an approximate relaxation mode given by
\begin{equation}
\bar{X}_{p}=
{\bar{\bm f}_{p}}{}^{\rm T} {\rm e}^{-\iGa^\dagger t_0/2}{\bm \Delta},
\label{XpDelta}
\end{equation}
where ${\bm \Delta}$, as a function of the state $Q$ of the system, is
defined by
\begin{equation}
	{\bm \Delta}(Q) =
	(\delta_1(Q), \delta_2(Q),\ldots, \delta_n(Q))^{\rm T}
\label{Deltadef}
\end{equation}
with
\begin{equation}
	\delta_i(Q)
	=
	\left\{
		\begin{array}{lcl}
			1 & {\rm for} & Q \in S_i,\\
			0 & {\rm for} & Q \notin S_i.
		\end{array}
	\right.
\end{equation}
The operator ${\rm e}^{-\iGa^\dagger t_0/2}$ reduces the contributions of faster modes in ${\bm \Delta}$.
Then, the variational problem, which is equivalent to
the eigenvalue problems (\ref{aeB1}) and (\ref{aeB2}) for
the conditional probability density $T_\tau(Q|Q')$ defined by (\ref{aeA11}),
leads to a generalized eigenvalue problem
\begin{equation}
\bar{C}(t_0+\tau)\bar{\bm f}_p
= {\rm e}^{-\bar{\lambda}_p \tau}
\bar{C}(t_0)\bar{\bm f}_p
\label{Cf}
\end{equation}
or
\begin{equation}
	\bar{\bm f}_p{}^{\rm T}
\bar{C}(t_0+\tau)
= {\rm e}^{-\bar{\lambda}_p \tau}
\bar{\bm f}_p{}^{\rm T}
\bar{C}(t_0)
\label{fTC}
\end{equation}
with
\begin{equation}
	\bar{\bm f}_p{}^{\rm T}
\bar{C}(t_0)
\bar{\bm f}_q
=\delta_{p,q},
\end{equation}
where
\begin{equation}
	\bar{C}(t)=
\langle {\bm \Delta}(t) {\bm \Delta}^{\rm T}(0) \rangle.
\end{equation}
Because $\bar{C}(t)=\bar{C}(t)^{\rm T}$,
Eqs.\ (\ref{Cf}) and (\ref{fTC}) are equivalent.
It follows from the definition of ${\bm \Delta}$ that
the $(i,j)$ component of $\bar{C}(t)$ is
the joint probability $\bar{P}_{i,j}(t)$:
$\bar{C}(t)=(\bar{P}_{i,j}(t))$.
\par
If we set $t_0 =0$, the generalized eigenvalue problem (\ref{fTC}) becomes
the eigenvalue problem (\ref{fTT}) with
$\bar{\Lambda}_p = {\rm e}^{-\bar{\lambda}_p \tau}$
or
$\bar{\tau}_p = 1/\bar{\lambda}_p$,
because
$\bar{C}(0)={\rm diag}(\bar{p}_1,\ldots,\bar{p}_n)$
and
$\bar{C}(\tau)\bar{C}(0)^{-1} = \bar{T}(\tau)$.
Thus, the Markov state model
is a special case of MSRMA with $t_0=0$.
Because
the operator
${\rm e}^{-\iGa^\dagger t_0/2}$
in Eq.\ (\ref{XpDelta})
reduces the contributions of faster modes in ${\bm \Delta}$,
the solutions of the generalized eigenvalue problem (\ref{Cf}) or (\ref{fTC})
become better approximations to the slow relaxation modes and rates
as $t_0$ becomes larger.
Therefore, the relaxation times $\bar{\tau}_p$ obtained by
the Markov state model are expected to be improved
by solving Eqs.\ (\ref{Cf}) or (\ref{fTC}) with $t_0 > 0$
rather than Eq.\ (\ref{fTT}).
\section{COMPUTATIONAL DETAILS}
An MD simulation was performed with the AMBER package (AMBER 11.0) with GPGPU
using the ff99SB force field and TIP3P model \cite{AMBER8}.
Chignolin consists of a 10-amino acid sequence: GYDPETGTWG.
We generated an extended structure of chignolin using the leap command and
solvated it with a 15 \AA \ buffer of TIP3P water around the peptide in each direction.
The numbers of atoms of chignolin and water molecules are 138 and 10,941 (3647 water molecules), respectively.
Two potassium ions (Na$^+$) are included in the system, resulting in a net-neutral system.
The total number of atoms in the system is 11,081.
After energy minimization and heating, equilibration at a constant pressure (1 atm) and 450 K,
a 50-ns MD simulation, was performed.
Finally, a 750-ns MD simulation was performed following the equilibration at 1 atm and 450 K.
We used a time step of 2 fs.
The Langevin thermostat
with a friction constant $\gamma =2.0~\mathrm{ps}^{-1}$ was used for temperature control.
The cutoff is 8 \AA, which was used to limit the direct space sum for the 
Particle Mesh Ewald (PME) method of AMBER.
For the equilibration and production run,
pmemd with GPGPU for MD simulations was used.
For analysis, the coordinates were saved every 10 ps.
The number of samples was 750,000.
\par
We used the coordinates of C$_{\alpha}$ atoms on the backbone as coordinates, so that the degrees of freedom was 30.
After removing the translational and rotational motions from the coordinates
of $C_{\alpha}$ atoms \cite{ROT,ROT2},
PCA and RMA were carried out on the coordinates of $C_{\alpha}$ atoms.
For RMA,
we set $t_0$ and $\tau$ to 10.0 ps and 20.0 ps, respectively.
\section{RESULTS AND DISCUSSION}
\subsection{Simulation Results}
We performed a 750-ns MD simulation of chignolin in aqueous solution at 450 K.
Before the present simulation, we performed two several-$\mu$s simulations of chignolin at 300 K.
In one simulation, chignolin folded to the native structure (see Fig.~\ref{Fig9}(a))
while in the other simulation, it folded to the misfolded structure (see Fig.~\ref{Fig9}(b)).
After folding to the native or misfolded structures once, these structures were maintained, with few transitions between the structures,
because the simulation times were not sufficiently long.
To allow for large conformational changes between the structures to occur frequently and to be able to observe numerous transitions during a several hundred-ns simulation,
we performed the simulation close to the transition temperature.
\par
\begin{figure}[ht]
\centerline{%
\includegraphics{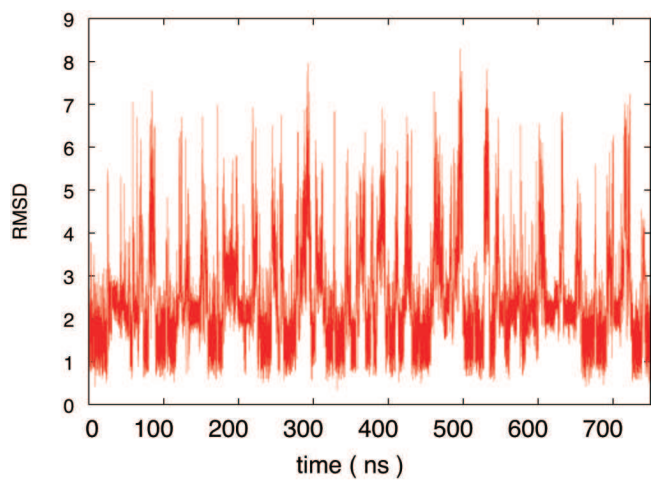}%
}
\caption{The time series of RMSD (\AA) from the native structure of chignolin near a transition temperature.
There are many transitions among
the native-like structures (RMSD $\thickapprox$ 1 \AA), misfolded structures (RMSD $\thickapprox$ 2 \AA),
and unfolded structures (RMSD $\thickapprox$ 5 \AA).}
\label{Fig1}
\end{figure}

The time series of the RMSD of $C_{\alpha}$ atoms
from the native structure
is shown in Fig.~\ref{Fig1}.
Here, the native structure is the first coordinate of 1UAO.pdb.
RMSD is calculated after fitting the obtained structures to the native structure.
Many transitions among the native-like structures (RMSD $\thickapprox$ 1 \AA), misfolded structures (RMSD $\thickapprox$ 2 \AA),
and unfolded structures (RMSD $\thickapprox$ 5 \AA)
occurred during the simulation.
\par
This system is suitable for testing the effectiveness of RMA, because of the many transitions observed among structures, including with the
completely unfolded structures.
The present simulation conducted at 450 K corresponds to a long simulation conducted at 300 K,
where many transitions are observed, similar to the observations in Ref.\ \citenum{Shaw}.
We next analyzed the dynamics of chignolin in the system using PCA, RMA, and the newly proposed combined method MSRMA.
\subsection{Results of principal component analysis}
\begin{figure}[ht]
\centerline{%
\includegraphics{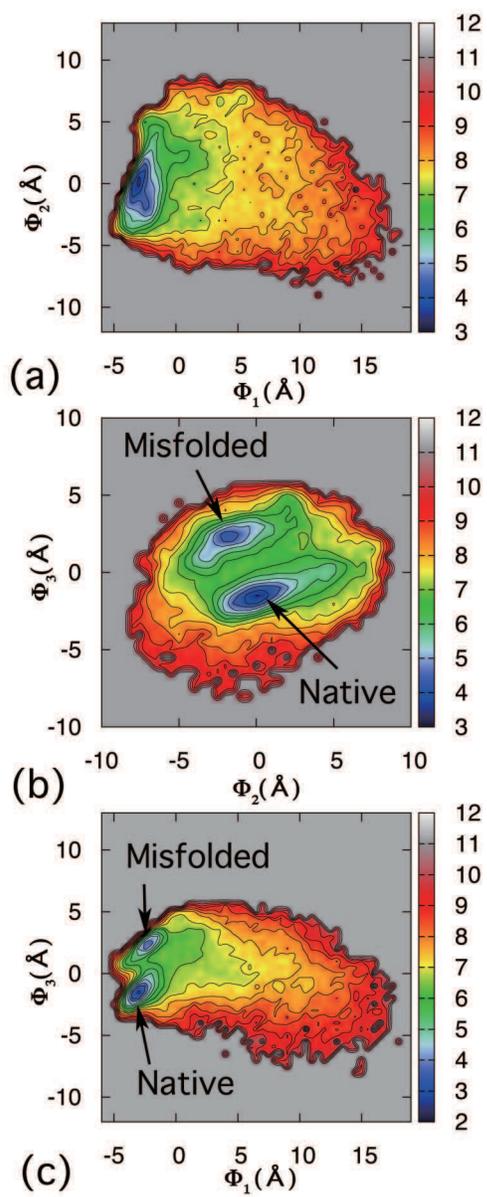}%
}
\caption{The free-energy surfaces of (a) $\Phi_1$ and $\Phi_2$, (b) $\Phi_2$ and $\Phi_3$, and (c) $\Phi_1$ and $\Phi_3$.}
\label{Fig2}
\end{figure}

\begin{figure}[ht]
\centerline{%
\includegraphics{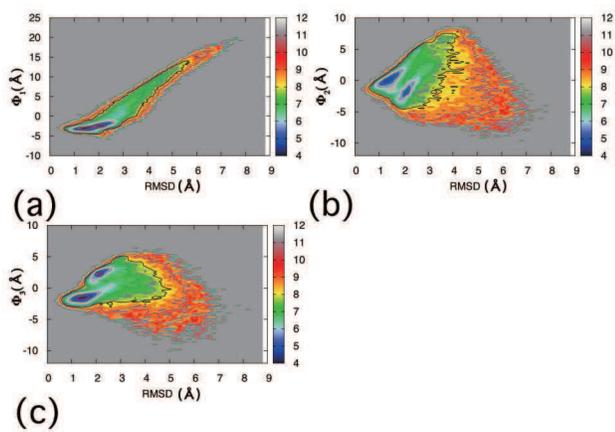}%
}
\caption{The relations between RMSD and (a) $\Phi_1$, (b) $\Phi_2$, or (c) $\Phi_3$.}
\label{Fig3}
\end{figure}

\begin{figure}[ht]
\centerline{%
\includegraphics[width=8cm]{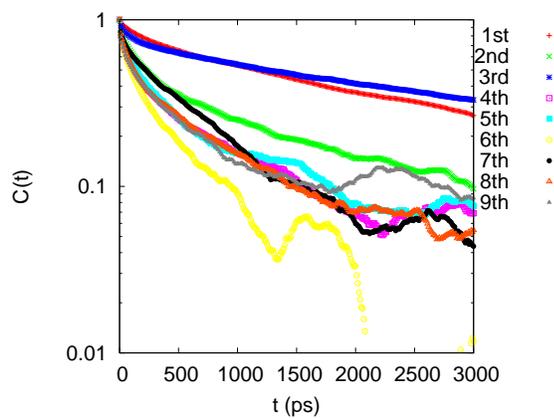}%
}
\caption{The time-displaced autocorrelation functions $\langle \Phi_p(t)\Phi_p(0) \rangle$ of $\Phi_p$ ($p=1,\cdots,9$) for PCA.
}
\label{Fig4}
\end{figure}

After a 750-ns production run was performed and the translational and rotational motions
were removed from the coordinates of $C_{\alpha}$ atoms \cite{ROT,ROT2}, we applied PCA to the system.
The five largest eigenvalues obtained by PCA are listed in Table I.
The $p$th eigenvalue is the variance of the $p$th principal component.
The first PC mode mainly contributes to the global conformational fluctuation around the average structure.
Figures~\ref{Fig2}(a) and \ref{Fig2}(b) show the free energy surfaces along
the first and second PC axes ($\Phi_1$ vs.\ $\Phi_2$), and the second and third PC axes
($\Phi_2$ vs.\ $\Phi_3$), respectively, given by Eq.\ (\ref{FreeEnergyPCA}).
We also show the free-energy surface of $\Phi_1$ and $\Phi_3$ in Fig.~\ref{Fig2}(c), because the two modes have a significantly slower relaxation than the others (see Fig.~\ref{Fig4} and its discussion).
The relations between RMSD from the native structure and PC components ((a) $\Phi_1$, (b) $\Phi_2$, or (c) $\Phi_3$) are shown in Fig.~\ref{Fig3}.
In Fig.~\ref{Fig3}(a), a local minimum-energy state can be observed near the value of $\Phi_1 \thickapprox -4$.
As the values of $\Phi_1$ become large, the values of RMSD also become large.
These results indicate that the direction of the first PC mode corresponds to the transition between the compact and unfolded structures.
From the free-energy surface of $\Phi_1$ and $\Phi_2$, we could not classify the native and misfolded states because the conformational difference between them is small compared with the conformational fluctuations of the system.
As shown in Fig.~\ref{Fig3}(b),
the direction of the second PC mode corresponds to a large fluctuation around
slightly compact structures (RMSD $\thickapprox$ 4), which does not correspond to the native and misfolded
structures.
Because of the two local minimum-energy states observed in Figs.~\ref{Fig2}(b) and
\ref{Fig2}(c) along the third PC axis,
the direction of the third PC mode corresponds to the transition between two minimum states, which
correspond to the native and misfolded states, even though there are many overlapping unfolded structures (see also Fig.~\ref{Fig10}).
The free-energy surface of $\Phi_1$ and $\Phi_3$ can classify not only
the native and misfolded states but also the unfolded state.
It is suggested that it is more effective to use PC modes with slower relaxation rather than those with larger conformational fluctuations as the axes of the free-energy surface.
However, PCA does not provide time information, and
the native and misfolded states still show slight overlap on the free-energy surface of $\Phi_1$ and $\Phi_3$ (see Fig.~\ref{Fig10}(c) and its discussion).
\par
Therefore, we next calculated the time-displaced autocorrelation functions of $\Phi_1$, $\Phi_2$, and $\Phi_3$, which are shown in Fig.~\ref{Fig4}.
The first and third PC modes (in red and blue, respectively) show slow relaxation, while the second PC mode (in green) shows relatively faster relaxation.
The second PC mode corresponds to large fluctuation of the slightly compact structures (RMSD $\thickapprox$ 4), which
are not completely compact like the native structure.
Therefore, PCA could extract the large conformational fluctuation around
the slightly compact structure as the second PC mode,
even though its relaxation is faster.
\begin{table}[htdp]
 \caption{The eigenvalues $\Lambda_p$ of the first to fifth PC modes, and
their relaxation times $\tau_p (=1/\lambda_p)$ and conformational fluctuations ${\tilde{\bm g}_p}^2$
of the first to fifth slowest relaxation modes.}
\begin{center}
\begin{tabular}{c||c||c|c}
 \hline
     & PCA & \multicolumn{2}{|c} {RMA} \\
\hline
 $p$ & $\Lambda_p$ (\AA$^2$) & $\tau_p$ (ps) & ${\tilde{\bm g}_p}^2$ (\AA$^2$)\\
\hline
1 &  20.17 &3.7 $\times 10^3$ & 6.93\\
2 &  7.01  &  2.4 $\times 10^3$ & 10.81\\
3 &  4.53  &  1.5 $\times 10^3$ & 1.85\\
4 &  2.58  &  7.5 $\times 10^2$ & 2.87\\
5 &  1.60  &  6.4 $\times 10^2$ & 1.74\\
\hline
\end{tabular}
\end{center}
\label{Tab1}
\end{table}
\subsection{Results of relaxation mode analysis}
We applied RMA to the same coordinate data
analyzed with PCA described above.
In Supplemental Figure S1 \cite{supple}, the time-displaced autocorrelation functions
of the $x$, $y$, $z$-coordinates for the $i$th C$_{\alpha}$ atom
obtained by the simulation directly and reproduced by RMA from Eq.\ (\ref{B13}) are compared.
These functions showed good agreement overall.
This means that
the appropriate relaxation modes and times were obtained.
\par
\begin{figure}[ht]
\centerline{%
\includegraphics{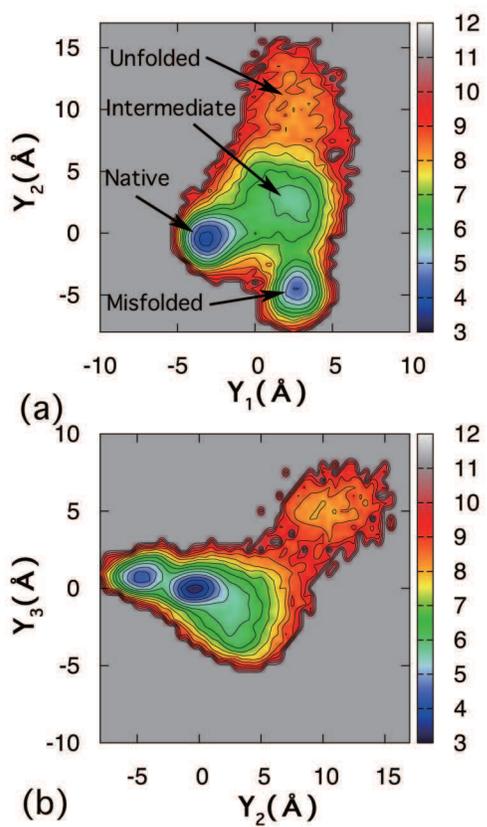}%
}
\caption{The free-energy surfaces of (a) $Y_1$ and $Y_2$, and (b) $Y_2$ and $Y_3$ for RMA}
\label{Fig5}
\end{figure}
\begin{figure}[ht]
\centerline{%
\includegraphics{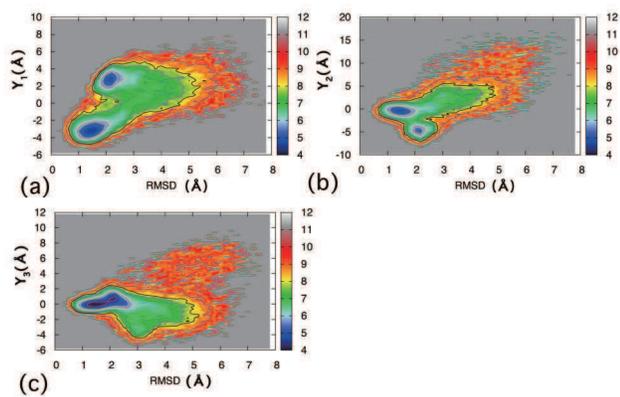}%
}
\caption{The relations between RMSD and (a) $Y_1$, (b) $Y_2$, or (c) $Y_3$(c).}
\label{Fig6}
\end{figure}

Table I shows the relaxation times and conformational fluctuations
of the five slowest relaxation modes (RMs).
The three slowest relaxation modes were slower than the other modes.
The second slowest relaxation mode showed the largest conformational fluctuation.
Figure~\ref{Fig5}(a) and \ref{Fig5}(b) show the free-energy surfaces along
the first and second slowest RM axes ($Y_1$ vs.\ $Y_2$) and
the second and third slowest RM axes ($Y_2$ vs.\ $Y_3$), respectively.
In Fig.~\ref{Fig6}, the relations between the RMSD from the native structure and RM components ($Y_1$(a), $Y_2$(b), or $Y_3$(c)) are shown.
In Fig.~\ref{Fig5}(a), there are four characteristic regions.
Two of them correspond to the local minimum-energy states of the native and misfolded structures.
Random structures are located at the larger values of $Y_2$ and $Y_3$.
There is a free-energy minimum near the origin of Fig.~\ref{Fig5}(a).
We refer to the state of the minimum as the intermediate state.
Figures~\ref{Fig5}(a) and \ref{Fig6}(a) indicate that the direction of the first RM corresponds to the transition between the native and misfolded states.
Figures~\ref{Fig5}(a) and \ref{Fig6}(b) indicate
that the direction of the second RM corresponds to the transition between the misfolded state and the intermediate state.
Furthermore, Figs.~\ref{Fig5}(b) and ~\ref{Fig6}(c) suggest that
the direction of the third RM may correspond to the transition between compact structures and the unfolded state.

\begin{figure}
\centerline{%
\includegraphics[width=8cm]{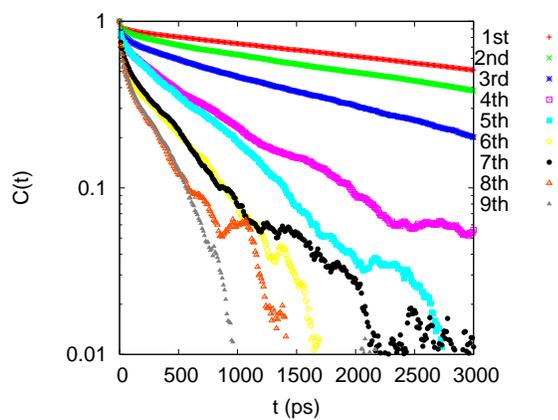}%
}
\caption{The time-displaced autocorrelation functions $\langle Y_p(t)Y_p(0) \rangle$ of $Y_p$, ($p=1,\cdots,9$) for RMA.
}
\label{Fig7}
\end{figure}

Note that the normalized $\{{\bm g}_p\}$ are not orthogonal to each other, in contrast to $\{{\bm F}_p\}$.
However, if the local minimum-energy states tend to be located in parallel along the slow RM axis,
the direction of the axis corresponds to
that for the transition between local minimum-energy states.
In the previous work of Ref.\ \citenum{MIT},
low free-energy paths, which connect to local minimum-energy states, were observed along the slow RM axes.
 The same tendency was observed in the present study.
Fig.~\ref{Fig7} shows the time-displaced autocorrelation functions
of the $p$th RM component $Y_p$.
The relaxation of $\langle Y_p(t) Y_p(0) \rangle$ becomes gradually faster as $p$ becomes larger.
Therefore, we succeeded in obtaining the order parameters with slow relaxation.
\par
\begin{figure}
\centerline{%
\includegraphics{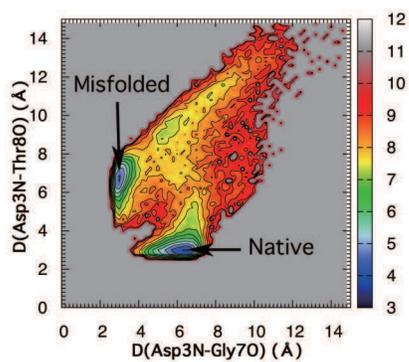}%
}
\caption{The free-energy surface along the distances between the amide nitrogen atom of Asp3 and the carbonyl oxygen atom of
Thr8, D(Asp3N-Thr8O), and
between the amide nitrogen atom of Asp3 and the carbonyl oxygen atom of Gly7,
D(Asp3N-Gly7O).}
\label{Fig8}
\end{figure}
\begin{figure}[ht]
\centerline{%
\includegraphics{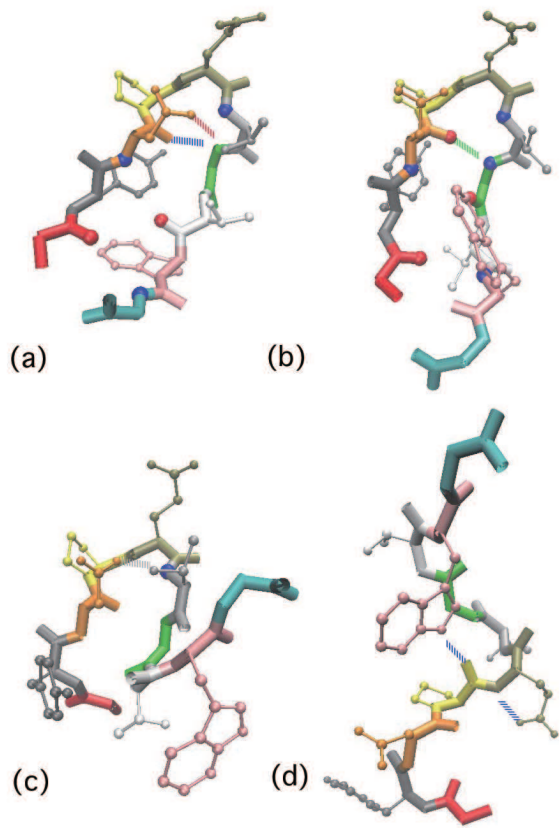}%
}
\caption{Structures of the (a) native state,
(b) misfolded state, (c) intermediate state, and (d) unfolded state (Multimedia view).
The structures are fitted on the backbone atoms from Asp3 to Glu5.
The movies of the snapshots are made using 
Visual Molecular Dynmics (VMD) \cite{vmd}.
The residues of Gly1 and Gly10 are shown in red and cyan, respectively.
Some characteristic hydrogen bonds obtained by the Hbonds plugin of VMD with
the cut-off distance of 3.0 \AA \ and the cut-off angle of 30 degrees are
also shown. Some important oxygen and nitrogen atoms of the backbone are shown as
red and blue spheres, respectively. }
\label{Fig9}
\end{figure}
\begin{figure}[ht]
\centerline{
\includegraphics{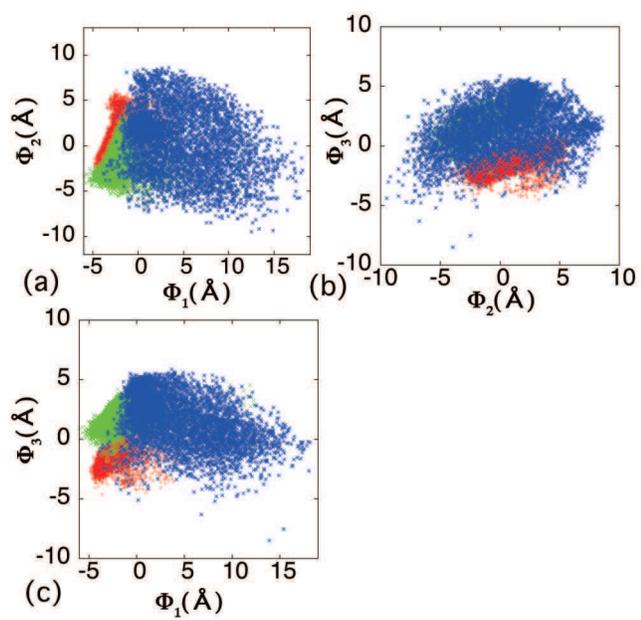}%
}
\caption{The distributions for the native state (red),
misfolded state (green), and intermediate state (blue)
in the planes of (a) $\Phi_1$ and $\Phi_2$, (b) $\Phi_2$ and $\Phi_3$, and
(c) $\Phi_1$ and $\Phi_3$.}
\label{Fig10}
\end{figure}

The suitable order parameters to identify the native and misfolded structures in the chignolin system have been identified in previous work, as follows:
the distance between the amide nitrogen atom of Asp3 (Asp3N) and the carbonyl oxygen atom of Gly7 (Gly7O),
D(Asp3N-Gly7O), and that between
Asp3N and the carbonyl oxygen atom of Thr8 (Thr8O), D(Asp3N-Thr8O) \cite{terada}.
These order parameters could clearly distinguish between the native and misfolded states.
Fig.~\ref{Fig8} shows the free-energy surface
along D(Asp3N-Gly7O) and D(Asp3N-Thr8O) obtained by the present simulation.
The shape of the free-energy surface is similar to that obtained by Refs.\ 
\citenum{terada} and \citenum{haradakitao}.
Although these order parameters are powerful for the system of chignolin,
we generally need to search for good order parameters that will
depend on the system.
From RMA, we obtained good order parameters automatically not only to identify the local minimum-energy states but also
to investigate the transitions between them.
\par
In addition, we obtained interesting results for the intermediate state from RMA.
The structures extracted for the four regions are listed in Table II,
which correspond to the native, misfolded, intermediate, and unfolded states.
The numbers of structures extracted for the native, misfolded, intermediate, and
unfolded states are 24,824, 14,571, 5251, and 1806, respectively, and the total number of samples was 75,000.
Fig.~\ref{Fig9} shows movies of the structures in the four clusters
(Multimedia view).
We extracted every 100 structures for the native and misfolded states,
every 20 structures for the intermediate state, and every 10 structures for the unfolded state to reduce the file size.
Each movie shows only a set of structures and not a time sequence; thus, the order of the structures has no meaning.
The native and misfolded structures appear as hairpin-like structures.
The native state is stabilized by four hydrogen bonds:
(a) between the oxygen atom of the side-chain carboxyl group of Asp3 (Asp3O$_\delta$) and
the amide nitrogen atom of Thr6 (Thr6N),
(b) between Asp3O$_\delta$ and the oxygen atom of the side-chain of Thr6 (Thr6O$_\gamma$),
(c) between Asp3N and Thr8O,
and (d) between the carobonyl oxygen atom of Gly1 (Gly1O) and
the amide nitrogen atom of GlY10 (Gly10N).
In addition, the side chains of Trp9 and Tyr2 are generally
located on the same side.
On the other hand,
the misfolded state is stabilized by five hydrogen bonds:
(a) between Asp3O$_\delta$ and Thr6N,
(b) between Asp3O$_\delta$ and Thr6O$_\gamma$,
(c) between the carbonyl oxygen atom of Asp3 (Asp3O) and
the amide nitrogen atom of Gly7 (Gly7N),
(d) between Asp3N and Gly7O,
and (e) between Gly1O and the amide nitrogen atom of Gly9 (Gly9N).
In addition, the side chains of Trp9 and Tyr2 are generally located on opposite sides.
The terminal residues show large fluctuations.
The misfolded state has different hydrogen bond patterns between Asp3N and Gly7O and between Gly1O and Gly9N.
The obtained native and misfolded structures were similar to those obtained in previous studies \cite{terada,taiji,haradakitao}.
\par
In the intermediate state, chignolin tends to form a turn from Asp3 to Glu5, similar to the native and misfolded states.
The intermediate state is stabilized by two hydrogen bonds:
(a) between Asp3O$_\delta$ and Thr6N,
and (b) between Asp3O$_\delta$ and Thr6O$_\gamma$.
These hydrogen bonds are also formed in the native and misfolded states.
Table III shows the average RMSD values of residues from Asp3 to Glu5 and from Tyr2 to Trp9 for the four clusters.
Here, we calculated the values using Visual Molecular Dynamics (VMD) \cite{vmd}.
The average RMSD values of residues from Asp3 to Glu5 of the native, misfolded, and intermediate structures were all small.
These results indicate that
the intermediate state has the characteristic structure,
i.e., the turn structure from Asp3 to Glu5.
\par
In Fig.\ S2 \cite{supple},
we show the Ramachandran plots of each residue from Tyr2 to Trp9 for the four clusters.
The plots of each residue from Tyr2 to Thr6 for the native and misfolded states are similar to each other.
The difference of the backbone dihedral angles of Gly7 causes the different hydrogen bond patterns observed between the native and misfolded states.
The plots of each residue from Asp3 to Glu5 for the intermediate state are similar to those for the native and misfolded states, indicating the formation of a turn structure.
These results demonstrate that the native, misfolded, and intermediate structures have the same turn structure.
The main difference between the unfolded state and the other states
is in the distribution of the dihedral angles of Pro4, as shown in Fig.\ S2 \cite{supple}.
This difference is responsible for the large RMSD value of residues from Asp3 to Glu5 of the unfolded state.
Based on the structures of the four states obtained by RMA,
we suggest that the dihedral angles are also good order parameters to
classify the states in this system.
\par
A previous report \cite{taiji} of a long simulation
suggested that chignolin folds to the native or misfolded structures through the turn structure present in the intermediate state.
The authors checked for the turn formation along their trajectories.
Fig.~S3 \cite{supple} shows the RMSD values from the native structure for the four states defined in Table III as a function of time. 
We observed the intermediate state in transition not only between the native and unfolded states
but also between the native and misfolded states.
Therefore, the slow modes extracted by RMA clarify the conformational transitions in the folding and
unfolding processes.
\par
Fig.~\ref{Fig10} shows the distributions for the native,
misfolded, and intermediate states classified on the free-energy surfaces obtained from RMA in the planes of (a) $\Phi_1$ and $\Phi_2$, (b) $\Phi_2$ and $\Phi_3$, and (c) $\Phi_1$ and $\Phi_3$.
Although the free energy of $\Phi_1$ and $\Phi_2$ cannot distinguish between the native and misfolded states, that of $\Phi_2$ and $\Phi_3$
could.
As shown in Fig.~\ref{Fig10}(c), there are still overlaps between the native and misfolded states.
The structures in the intermediate state are distributed widely on the free-energy surface obtained by PCA and overlap on the areas of the folded and misfolded states corresponding to the free-energy surface of $\Phi_2$ and $\Phi_3$.
PCA could not identify the intermediate state, because the structural fluctuation near
the terminal residues of the intermediate state is large and PCA extracts large
structural fluctuations.
RMA could extract the characteristic structure, i.e., the turn structure
from Asp3 to Glu5, because the large fluctuations of the terminal residues have
fast relaxation, and RMA neglects the faster conformational changes.
\par
\begin{figure}[ht]
\centerline{%
\includegraphics{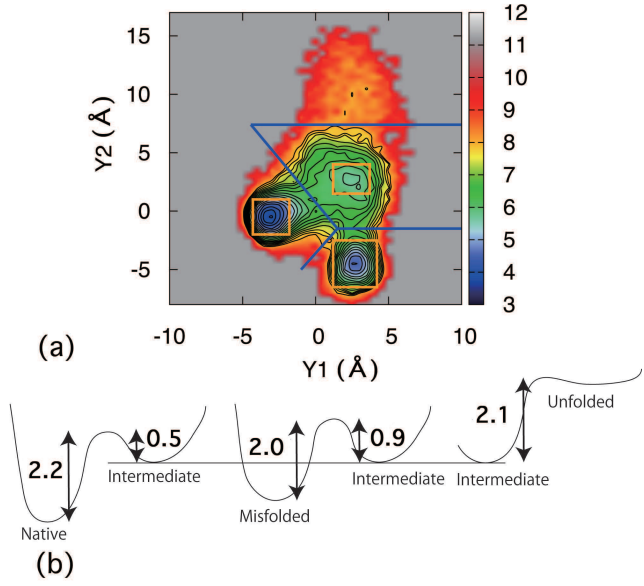}
}
\caption{
(a) The free-energy surface of $Y_1$ and $Y_2$ showing three regions for the native, misfolded, and intermediate structures listed in Table II (in orange) and the dividing lines for the four states of the Markov state models and the
Markov state relaxation mode analysis model (in blue).
(b) Schematic drawing of the free-energy differences between the intermediate stae and the other three states (native, misfolded, and unfolded states).
The contour lines are drawn in the range from 3 to 8 at an interval of 0.25.
The values in (b) are $RT \times \Delta F$,
where $\Delta F$ is the difference of the free energy given by Eq.\ (\ref{freerma}) and $RT$ is 0.89 kcal/mol at $T$ = 450 K.
The free energy differences are
approximately estimated by counting the numbers of contour lines.
}
\label{Fig11}
\end{figure}

These results strongly support the utility in clustering structures using the free energy obtained from RMA.
We next investigated the transitions between the four clusters in detail.
Fig.~\ref{Fig11}(a) shows the free energy surface of $Y_1$ and $Y_2$ with more contour lines.
The free energy differences between the intermediate and the other three states are shown schematically in Fig.~\ref{Fig11}(b).
The free energy difference from the intermediate state to the unfolded state
was calculated to be approximately 2.1 kcal/mol.
The free energy barrier seems to be small from the unfolded state to the transition state
between the unfolded and intermediate states.
Furthermore, given that the free energy difference from the intermediate state to
the transition state toward the native state
(0.5 kcal/mol)
is lower than that toward the misfolded state
(0.9 kcal/mol),
the transition from the intermediate state to the native state appears to be easier than that to the misfolded state.
These results suggest that the main folding path of chignolin from the unfolded to the native states
passes through the intermediate state.
The free energy difference from the native or misfolded states to their
transition states toward the intermediate state were both high (2.2 or 2.0 kcal/mol, respectively).
One reason could be that the hydrogen bonds between backbone atoms need to be broken in order to change from
the native or misfolded states to the intermediate state.
The slowest relaxation process is mostly related to the transition between the native and misfolded states; it corresponds to escape from the native state and the misfolded state to intermediate states.
The second slowest relaxation process seems to be related to the transition between misfolded and unfolded states; it corresponds to escape from the
misfolded state to the intermediate state and from the intermediate state to the unfolded state.


\subsection{Results of Markov state relaxation mode analysis}
Recently, Markov state models have been constructed in the discrete state space defined by the clustering in subspace determined by tICA \cite{MSM2,MSMPANDE}.
In the present work, we classified the structures into a smaller number of states by using the free-energy surface obtained by RMA, and then
applied the Markov state model and MSRMA to analyze the states.
\par
We divided the free-energy surface of $Y_1$ and $Y_2$ to the four regions shown in Fig.~\ref{Fig11}(a),
and classified the structures into the following four states:
native ($S_1$), misfolded ($S_2$), intermediate ($S_3$), and unfolded ($S_4$) states.
After calculating the trajectories of $\delta_i$ ($i=$1,$\cdots$,4), we solved the generalized eigenvalue problem of Eq.\ (\ref{Cf}).
Because ${\bar C(t)}$ is a symmetric matrix, ${\bar C(t)}={\bar C(t)}^{\rm T}$,
we used $\frac{1}{2} ({\bar C(t)}+{\bar C(t)}^T)$ instead of ${\bar C(t)}$.
\par
\begin{figure}[ht]
\centerline{%
\includegraphics{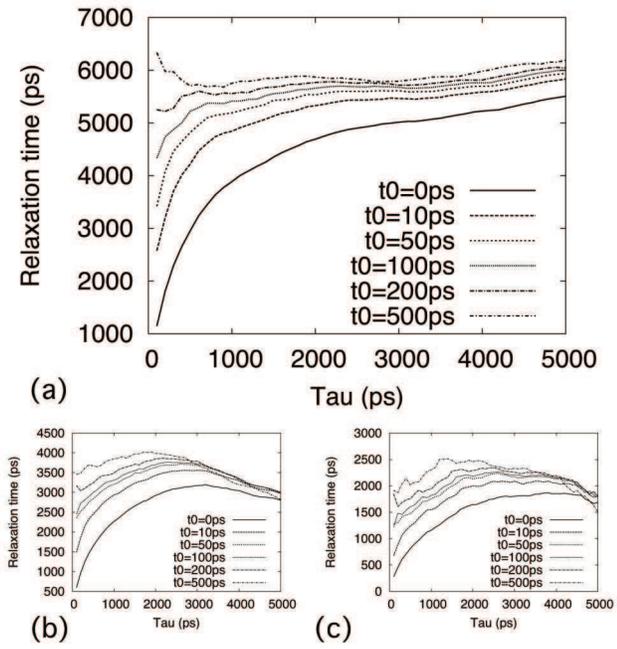}%
}
\caption{The relaxation times of the (a) second, (b) third, and (c) fourth relaxation modes obtained by MSRMA as a function of time interval $\tau$.
The lines of $t_0=0$ ps correspond to the results of a simple Markov state model. }
\label{Fig12}
\end{figure}

Fig.~\ref{Fig12} shows the relaxation times $\bar{\tau}_p=1/\bar{\lambda}_p$ obtained by MSRMA as a function of $\tau$ when $t_0=$0, 10, 50, 100, 200, and 500 ps.
Because the first eigenvector
$\bar{\bm f}_1$, proportional to $ (1, 1, 1, 1)^{\rm T}$,
corresponds to the steady state with infinite relaxation time $\tau_1 = \infty$,
we show the second (a), third (b), and fourth (c) longest relaxation times in Fig.~\ref{Fig12}.
The lines of $t_0=0$ correspond to the results of a simple Markov state model.
In the case of $t_0=0$, ${\bar \tau}_p$ ($p=$2, 3, and 4) values slowly approach the appropriate time scale, i.e.,
the values for plateau regions or peak values of the solid lines of Fig.~\ref{Fig12}, when $\tau$ is increased.
Using the method for determining the value of $\tau$ described in Section IID,
the appropriate time interval for $\tau$ was determined to be around 3 ns, which
corresponds to the time scale of
the slower relaxation times obtained by RMA.
The relaxation times at $\tau=3$ ns for the second, third, and fourth slowest relaxation modes were approximately 5, 3, and 1.8 ns, respectively, which
are close to but slightly higher than those obtained by RMA at 3.7, 2.4, and 1.7 ns, respectively.
\par
For the lines of $t_0 > 0$,
the values of ${\bar \tau}_p$ quickly approach the appropriate time scale; i.e.,
those corresponding to the values for plateau regions or peak values of Fig.~\ref{Fig12}.
Therefore, the slow relaxation times can be improved when applying MSRMA with $t_0 > 0$, which is introduced in order to
reduce the relative weight of the faster modes contained in the physical quantities given by Eq.\ (\ref{XpDelta}).
The relaxation times are improved even if $t_0$ is small; in particular,
the appropriate relaxation times are obtained
using a shorter value for the time interval $\tau$.
The estimated relaxation times were approximately 5.5 ns, 3.5 ns, and 2.0 ns
(using $t_0= 0.2$ ns and $\tau=1$ ns).
\par
\begin{figure}[ht]
\centerline{%
\includegraphics{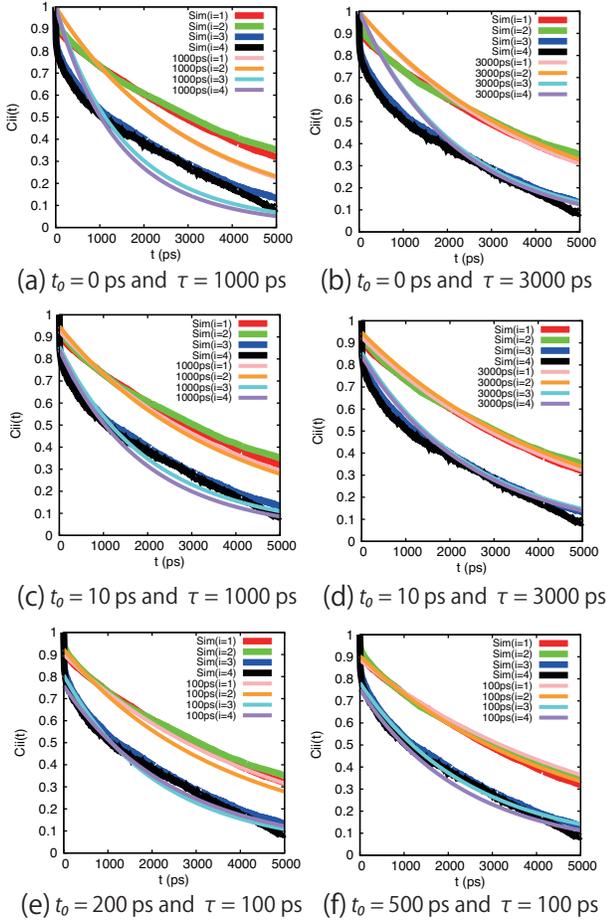}%
}
\caption{The normalized time-displaced autocorrelation functions
${\hat {\bar C}}_{i,i}(t)$ of Eq.\ (\ref{MScorr}) calculated directly (Sim) and
reproduced by Eq.\ (\ref{B13}). The values of $t_0=0$ and $\tau$ in picoseconds are (a) 0 and 1000, (b) 0 and 3000, (c) 10 and 1000, (d) 10 and 3000,
(e) 200 and 100, and (f) 500 and 100, respectively.
The results of $t_0=0$ ps correspond to those of the simple Markov state model.
}
\label{Fig13}
\end{figure}

The normalized time-displaced autocorrelation functions of ${\bm \Delta}$
both calculated directly and reproduced by MSRMA using Eq.\ (\ref{B13}) are shown in Fig.~\ref{Fig13}.
The function is given by
\begin{equation}
\begin{array}{ll}
{\hat {\bar C}}_{i,i}(t)&= \displaystyle \frac {\langle (\delta_i (t) - \langle \delta_i (t) \rangle)(\delta_i (0) - \langle \delta_i (0) \rangle) \rangle}
 {\langle (\delta_i (0) - \langle \delta_i (0) \rangle)^2 \rangle}\\
&\displaystyle =\frac{ \langle \delta_i (t) \delta_i (0) \rangle - {\bar p}(i)^2}{ {\bar p}(i)-{\bar p}(i)^2}~.
\end{array}
\label{MScorr}
\end{equation}
Note that,
in the present case,
the summation in Eq.\ (\ref{B13}) is taken from $p=$1 to 4.
The results obtained directly shown in Fig.~\ref{Fig13} indicate that
the time-displaced autocorrelation functions of $i=$ 1 and 2, which correspond to the native and misfolded states,
respectively, have similar slower relaxations.
The time-displaced autocorrelation function of $i=$3 and 4, which correspond to the intermediate and unfolded states, respectively, have similar relaxations, which are slightly faster compared to those of $i=$1 and 2.
\par
The correlation matrix $\bar{C}(t)$ reproduced by Eq.\ (\ref{B13})
must be equal to that directly calculated by the simulation
at $t=t_0$ and $t = t_0+\tau$,
because Eq.\ (\ref{B11b}) holds exactly for $t=0$ and $t=\tau$,
as mentioned above in Section \ref{RMA}.
Therefore, as shown in Fig.\ \ref{Fig13},
all lines of ${\hat {\bar C}}_{i,i}(t)$ reproduced by Eq.\ (\ref{B13})
go through the points of directly calculated
${\hat {\bar C}}_{i,i}(t)$
at $t=t_0$ and $t=t_0+\tau$.
In the case of $t_0=0$~ps and $\tau=1000$~ps, shown in Fig.~\ref{Fig13} (a),
reproduced ${\hat {\bar C}}_{i,i}(t)$ is larger than the directly calculated
${\hat {\bar C}}_{i,i}(t)$ for $t_0 = 0 < t < t_0 + \tau = \tau$,
and is smaller for  $t_0 + \tau = \tau < t$.
As a result, the relaxation times obtained by MSRMA are underestimated.
This is due to the above-mentioned restriction
and the existence of fast relaxation modes in
the directly calculated ${\hat {\bar C}}_{i,i}(t)$,
which causes the fast initial decay of ${\hat {\bar C}}_{i,i}(t)$.
These fast relaxation modes cannot be
described by the three relaxation modes used in MSRMA.
Thus, the results of MSRMA with $t_0=0$ are improved
by using a longer $\tau$ as shown in Fig.~\ref{Fig13}(b),
which is the usual method for the simple Markov state model.
Comparison of the results of MSRMA
for $t_0=10$ ps and $\tau = 1000$ ps
shown in Fig.\ \ref{Fig13}(c)
with those
for $t_0=0$ ps and $\tau = 1000$ ps
shown in Fig.\ \ref{Fig13}(a),
and
comparison of
those
for $t_0=10$ ps and $\tau = 3000$ ps
shown in Fig.\ \ref{Fig13}(d)
with those
for $t_0=0$ ps and $\tau = 3000$ ps
shown in Fig.\ \ref{Fig13}(b)
make it clear that
the results of MSRMA are improved by using finite $t_0$,
even if $t_0$ is small.
Fig.\ \ref{Fig12} clearly shows
that
the relaxation times obtained
for $t_0=10$ ps and $\tau = 1000$ ps
are similar to those obtained
for $t_0=0$ ps and $\tau = 3000$ ps.
The results
for $t_0=200$ ps and $\tau = 100$ ps
shown in Fig.\ \ref{Fig13}(e)
and
those
for $t_0=500$ ps and $\tau = 100$ ps
shown in Fig.\ \ref{Fig13}(f)
demonstrate that
a good description of the relaxation of the states
can be obtained by using MSRMA with finite $t_0$, even if $\tau$ is small.
Note that
the values of $t_0$ and $t_0 + \tau$
[200 ps and 300 ps for
Fig.\ \ref{Fig13}(e),
and
500 ps and 600 ps for
Fig.\ \ref{Fig13}(e)]
are much smaller than
the value of $\tau$ (3000 ps) used to estimate the relaxation times
by MSRMA with $t_0=0$, i.e., the simple Markov state model.
\begin{table}[htdp]
 \caption{
The definitions of four clusters: native state (Native),
misfolded state (Misfolded),
intermediate state (Intermediate), and unfolded state (Unfolded).
The total number of samples in the four regions is 46,452, and the total number of samples overall is 75,000.
The structures in the ranges from min to max of $Y_1$ and $Y_2$
(for native, misfolded, and intermediate states) or $Y_2$ and $Y_3$ (for unfolded state) were extracted.}
 \begin{center}
 \begin{tabular}{c||c||c||c||c}
 \hline
Cluster & Native & Misfolded & Intermediate & Unfolded\\
\hline
&$min$:$max$&$min$:$max$&$min$:$max$&$min$:$max$\\
\hline
$Y_1$& $-$4.3:$-$1.8& 1.2:4.2 & 1.2:3.7 &   \\
$Y_2$& $-$2.0:1.0 & $-$6.5:$-$2.5 & 1.5:4.0 & 5.0:16.0 \\
$Y_3$&          &           &         & 4.0:10.0 \\
\hline
No. of samples& 24824 & 14571 & 5251 & 1806\\
\hline
\end{tabular}
\end{center}
 \label{Tab3}
\end{table}

\begin{table}[htdp]
 \caption{The average RMSD values (\AA) for the four clusters.
The structures in all clusters were fit to the first structure in the native state using backbone atoms from Asp3 to Glu5.
Then, the average RMSD from the average structure of each cluster
was calculated for the backbone atoms
from Tyr2 to Trp9 or
from Asp3 to Glu5.
}
\begin{center}
\begin{tabular}{r|c|c}
 \hline
     & Tyr2 to Trp9& Asp3 to Glu5\\
\hline
Native &  1.198 & 0.185 \\
Misfolded &  1.652 & 0.212\\
Intermediate &  3.671 & 0.245\\
Unfolded &  8.406 & 0.993\\
\hline
\end{tabular}
\end{center}
\label{Tab5}
\end{table}

\section{CONCLUSIONS}
In this study, we applied RMA to extract reaction coordinates for a protein system characterized by large conformational changes such as folding/unfolding simulation.
We performed a 750-ns simulation of chignolin near its transition temperature and observed many
transitions between the most stable, misfolded, and unfolded states.
RMA could extract the slow relaxation modes.
The local minimum-energy states were usually stable and
the system remained in this state for a long time during the simulation.
The order parameters with slow relaxation corresponded to the directions between the local minimum-energy states.
Therefore, these slow relaxation modes indicate the suitable order parameters for identifying the local minimum-energy states
automatically.
From the free-energy surfaces obtained by RMA, we identified not only the native and misfolded states but also the intermediate state.
RMA neglected the conformational changes with fast relaxation and extracted the intermediate state showing a turn structure from Asp3 to Glu5.
Thus, using RMA clarified the folding process of chignolin to the native or misfolded structures through the turn structure.
Analysis of the free energy differences further revealed that the transition from the folding to native state through the turn structure is easier than that to the misfolded state.

Note that in simulations, the free-energy surface depends on both the temperature and simulation time in practice.
In general, it is still difficult to determine whether or a not a simulation time is sufficient.
The free-energy surface obtained in the present study resulted from a 750-ns simulation near the transition temperature.
To support the validity of the free-energy surface, we also need to investigate
the stability of the local minimum-energy states.
In addition, the free-energy surface near a transition temperature is different
from that at a room temperature.
Nevertheless, understanding the dynamics and kinetics of chignolin near its transition temperature can provide important information on the folding process.
Overall, these results indicated that RMA can be used to effectively analyze long simulations at room temperature and is also useful for investigating systems with large conformational changes, such as intrinsically disordered proteins and protein folding.
\par
Our experience of RMA suggests that a simulation time longer than approximately 10 $\tau_{\rm slow}$ is preferable, where $\tau_{\rm slow}$ is the slowest relaxation time. In addition, even if such a long simulation cannot be performed, rare transitions, (e.g., one transition), can be obtained during the simulation
to identify local minimum-energy states and the transitions between them.
\par
We have here proposed a new analysis method called MSRMA.
By applying RMA to the Markov state model,
we introduced the evolution time $t_0$,
which reduced the relative weight of the faster modes contained in the physical quantities.
From clustering the states using the free-energy surface obtained by RMA, we constructed a Markov state model and
performed
MSRMA.
Analysis and comparison of the relaxation times of the states obtained by a simple Markov state model and MSRMA showed
similar relaxation times to those obtained from RMA.
Moreover, we show that MSRMA clearly improves the approximate relaxation times.

\begin{acknowledgments}
This work was supported, in part, by the Grant-in-Aid for Scientific Research on Innovative Areas ``Material Design through Computics'' from the Ministry of Education, Culture, Sports, Science and Technology (MEXT), Japan.
This research is partially supported by Initiative on Promotion of Supercomputing for Young or Women Researchers, Supercomputing Division, Information Technology Center, The University of Tokyo, and  HA-PACS at the CCS, University of Tsukuba.
\end{acknowledgments}
\appendix
\section{Relaxation modes}
\label{appendixA}
We consider a Langevin equation for a biomolecule with $N$ atoms:
\begin{equation}
\dis m_i\frac{d \bm{v}_i}{dt} = -\zeta \bm{v}_i
-\frac{\partial}{\partial \bm{r}_i} U( \{ \bm{r}_j \})+\bm{w}_i
\label{aeA1}
\end{equation}
with
\begin{equation}
\frac{d \bm{r}_i}{dt}=\bm{v}_i~.
\label{aeA2}
\end{equation}
Here, $\bm{r}_i(t)$ and $\bm{v}_i(t)$ denote the position and the velocity of the $i$th atom at time $t$, respectively.
$m_i$ is the mass of the $i$th atom and $\zeta$ is the friction constant.
The interaction between atoms is described by
the potential $U(\{\bm{r}_i\})=U(\bm{r}_1,\ldots,\bm{r}_N)$.
The random force $\bm{w}_i(t)$ acting on the $i$th atom is a Gaussian white stochastic process and satisfies
\begin{equation}
\langle w_{i,\alpha}(t) w_{j,\beta}(t) \rangle =2 \zeta k_B T \delta_{\alpha,\beta}\delta_{i,j}\delta(t-t^{\prime}),
\label{aeA3}
\end{equation}
where $w_{i,\alpha}$, $k_B$, and $T$ denote the $\alpha$-component of $\bm{w}_i$ ($\alpha$=x, y, or z), the Boltzmann constant, and the temperature of the system, respectively.
The Kramers equation equivalent to Eqs.~(\ref{aeA1}) and (\ref{aeA2}) can be written as
\begin{equation}
\dis \frac{\partial}{\partial t}P(Q,t)=
-\iGa(Q) P(Q,t)~.
\label{aeA4}
\end{equation}
Here, $Q=\{\bm{r}_1,\ldots,\bm{r}_N,\bm{v}_1,\ldots,\bm{v}_N\}$ denotes
a point in the phase space of the system, and $P(Q,t)dQ$ denotes the probability
that the system is found at time $t$ in an infinitesimal volume $dQ$ at the
point $Q$ in the phase space.
The time-evolution operator $\iGa$ is explicitly given by
\begin{equation}
\begin{array}{l}
\iGa(Q)= \\
\dis \sum_{i=1}^{N} \dis
\left\{
\frac{\partial}{\partial \bm{r}_i} \cdot \bm{v}_i
-\frac{1}{m_i}\frac{\partial}{\partial \bm{v}_i} \cdot
\frac{\partial U}{\partial \bm{r}_i}
- \frac{\zeta}{m_i}\frac{\partial}{\partial \bm{v}_i}\cdot
\left({\bm v}_i + \frac{k_B T}{m_i} \frac{\partial}{\partial \bm{v}_i}
\right)
\right\}~.
\end{array}
\label{aeA5}
\end{equation}
Because Eqs.~(\ref{aeA1}) and  (\ref{aeA2}) are nonlinear for anharmonic potentials,
it is generally difficult to
define relaxation modes as normal coordinates of the equations.
However, as explained below, we can define relaxation modes on the
basis of Eq.\ (\ref{aeA4}), which is a linear equation.
\par
Let $\phi_n(Q)$ and $\psi_n(Q)$ denote the left and right eigenvectors of
the time-evolution operator $\iGa(Q)$ with an eigenvalue $\lambda_n$, respectively:
\begin{equation}
\iGa(Q) \psi_n(Q)
=\lambda_n\psi_n(Q) ~\\
\label{aeA6a}
\end{equation}
and
\begin{equation}
\iGa^{\dagger}(Q)\phi_n(Q)
=\lambda_n\phi_n(Q).
\label{aeA6b}
\end{equation}
Here, $\iGa^{\dagger}(Q)$ is the adjoint operator of $\iGa(Q)$ defined by
\begin{equation}
\int f(Q)(\iGa(Q)g(Q)) dQ=\int(\iGa^{\dagger}(Q)f(Q))g(Q)dQ~,
\label{aeA7}
\end{equation}
which satisfies
\begin{equation}
\iGa(Q)\delta(Q-Q')=\iGa^{\dagger}(Q')\delta(Q-Q')~.
\end{equation}
The eigenvectors are chosen to satisfy the orthonormal condition:
\begin{equation}
\int dQ \phi_n(Q)\psi_m(Q)=\delta_{n,m}~.
\label{aeA8}
\end{equation}
The equilibrium distribution function for Kramers equation
\begin{equation}
P_{\rm eq}(Q) \dis \propto \exp  \left[-\frac{1}{k_BT} \left\{ \frac{1}{2}
\sum_{i} m_i \bm{v}_i^2 + U (\{\bm{r}_j\}) \right\} \right]
\label{aeA9}
\end{equation}
is a right eigenfunction of $\iGa(Q)$ with zero eigenvalue: $\iGa(Q) P_{\rm eq} (Q) =0$.
The corresponding left eigenfunction is 1.
\par
From Eq.\ (\ref{aeA4}), a formal solution of $P(Q,t)$ for an initial condition $P(Q, 0)=P_0(Q)$ is given by
\begin{equation}
P(Q,t)=e^{-\iGa(Q) t}P_0(Q)~.
\label{aeA10}
\end{equation}
The conditional probability $T_t(Q|Q')dQ$ that
the system is found at time $t$ in an infinitesimal volume $dQ$ at $Q$,
given that the system is at $Q'$ at time $0$, is then given by
\begin{equation}
	T_t(Q|Q')={ \rm e}^{-\iGa(Q) t}\delta(Q-Q')~.
\label{aeA11}
\end{equation}
The time-displaced correlation functions of two physical quantities $A(Q)$ and $B(Q)$
in the equilibrium are given by
\begin{align}
	\langle A(t)B(0) \rangle
	& = \int \int A(Q) T_t(Q|Q')B(Q')P_{\rm eq}(Q') dQ dQ'
	\nonumber\\
	& = \int A(Q) {\rm e}^{ -\iGa(Q)t}B(Q)P_{\rm eq}(Q) dQ
	\nonumber\\
	& =\int \left\{{\rm e}^{-\iGa^{\dagger}(Q) t}A(Q) \right\}B(Q)P_{\rm eq}(Q) dQ
\label{aeA12}
\end{align}
We define a quantity $\hat{\phi}_n(Q)$ through
\begin{equation}
\psi_n(Q)=\hat{\phi}_n(Q)P_{\rm eq}(Q)~.
\label{aeA13}
\end{equation}
From Eq.\ (\ref{aeA12}),
the equilibrium time-displaced correlation function of $\phi_n(Q)$ and $\hat{\phi}_m(Q)$
is given by
\begin{equation}
	\langle \phi_n(t) \hat{\phi}_m(0) \rangle=\delta_{n,m} {\rm e}^{-\lambda_n t}~.
\label{aeA14}
\end{equation}
If two quantities $A(Q)$ and $B(Q)$
are expanded as
\begin{equation}
A(Q) =\sum_n a_n \phi_n (Q)~~\mathrm{ and }~~ B(Q) =\sum_n \hat{b}_n \hat{\phi}_n (Q),
\label{aeA15}
\end{equation}
with
\begin{equation}
a_n=\int dQ A(Q)\hat{\phi}_n(Q)P_{\rm eq}(Q)=\langle A\hat{\phi}_n \rangle
\label{aeA16}
\end{equation}
and
\begin{equation}
\hat{b}_n=\int dQ \phi_n(Q) B(Q) P_{\rm eq}(Q)=\langle \phi_n B \rangle,
\label{aeA17}
\end{equation}
then the time correlation function of $A$ and $B$ in the equilibrium is given by
\begin{equation}
\langle A(t) B(0) \rangle =\sum_n a_n \hat{b}_n e^{-\lambda_n t}~.
\label{aeA18}
\end{equation}
Thus, in terms of $\phi_n(Q)$ and $\hat{\phi}_n(Q)$, the correlation
function $\langle A(t) B(0) \rangle$ is decomposed into a sum of exponentially relaxing contributions.
Therefore, we use two sets of functions, $\{\phi_n(Q) \}$ and $\{ \hat{\phi}_n(Q) \}$, as relaxation modes and call $\{\lambda_n\}$ their relaxation rates.
Note that the eigenvalues $\lambda_n$ are not necessarily real in general.
Therefore, the terms ${\rm e}^{-\lambda_n t}$ in Eq.\ (\ref{aeA18}) can describe oscillatory behavior \cite{HKT}.
\par
The Kramers equation satisfies
the detailed balance condition \cite{R}
\begin{equation}
P_{\rm eq}(Q')\iGa(Q) \delta(Q-Q')=
P_{\rm eq}(\epsilon Q)\iGa^{\dagger}(\epsilon Q) \delta(\epsilon Q- \epsilon Q')~,
\label{db}
\end{equation}
and
\begin{equation}
P_{\rm eq}(Q)=P_{\rm eq}(\epsilon Q)~,
\end{equation}
where $\epsilon Q$ denotes the time-reversed state of the state $Q$,
namely $\epsilon Q=\{\epsilon_1 \bm{r}_1,\ldots,\epsilon_N \bm{r}_N,
. \epsilon_{N+1} \bm{v}_1,\ldots,\epsilon_{2N} \bm{v}_N\}$ with
\begin{equation}
\epsilon_i =
\left\{
	\begin{array}{rcl}
		1 & {\rm for} & i=1,\ldots,N,\\
		-1& {\rm for} & i=N+1,\ldots,2N.
	\end{array}
\right.
\label{aeA19}
\end{equation}
This leads to the relation
\begin{equation}
\hat{\phi}_n(Q) \propto \phi_n(\epsilon Q)
\label{aeA20}
\end{equation}
between the relaxation modes.
\section{Relaxation mode analysis}
The eigenvalue problem of eqs. (\ref{aeA6a}) and (\ref{aeA6b}) can be written as
\begin{equation}
e^{-\iGa(Q) \tau} \hat{\phi}_n(Q) P_{\rm eq}(Q)=
e^{-\lambda_n \tau} \hat{\phi}_n(Q) P_{\rm eq}(Q)
\label{aeB1}
\end{equation}
and
\begin{equation}
e^{-\iGa^{\dagger}(Q) \tau} \phi_n(Q) =
e^{-\lambda_n \tau} \phi_n(Q)~.
\label{aeB2}
\end{equation}
This eigenvalue problem is equivalent to
the variational problem
\begin{equation}
\delta {\cal R}=0
\label{aeB3}
\end{equation}
for
\begin{equation}
\dis \delta {\cal R}[\phi_n,\hat{\phi}_n] =
\frac{\langle \phi_n(\tau)\hat{\phi}_n(0) \rangle}
{\langle \phi_n\hat{\phi}_n\rangle}
\label{aeB4}
\end{equation}
and the stationary value of ${\cal R}$ gives the
eigenvalue $\exp(-\lambda_n \tau)$.
In Ref.\ \citenum{HKT}, a linear combination of relevant quantities is
used as a trial function for $\phi_n$ and $\hat{\phi}_n$,
and the variational problem is solved.
The solution is given as a solution of a generalized eigenvalue problem
for the equilibrium time-displaced correlation matrices of the quantities.
\par
We use the following functions as an approximate relaxation mode:
\begin{equation}
X_{p}=
{{\bm f}_{p}}^{\rm T} {\rm e}^{-\iGa^\dagger t_0/2}{\bm R}
\label{aeB5}
\end{equation}
and
\begin{equation}
\hat{X}_{p}=
P_{\rm eq}^{-1}
\hat{{\bm f}}^{\rm T}_{p}
{\rm e}^{-\iGa t_0/2}{\bm R} P_{\rm eq}
\label{aeB6}
\end{equation}
as the $p$th trial functions for $\phi_n(Q)$ and $\hat{\phi}_n(Q)$, respectively,
where
\begin{equation}
{\bm f}_p = (f_{p,1}, f_{p,2}, \ldots, f_{p,6N})^{\rm T}
\end{equation}
and
\begin{equation}
\hat{{\bm f}}_p = (\hat{f}_{p,1}, \hat{f}_{p,2}, \ldots, \hat{f}_{p,6N})^{\rm T}
\end{equation}
are variational parameters.
Here,
${\bm R}$ is defined by
\begin{equation}
{\bm R}=( {{\bm r}_1^{\prime}}^ {\rm T},{{\bm r}_2^{\prime}}^{\rm T},\ldots,{{\bm r}_N^{\prime}}^{\rm T}, {{\bm v}_1^{\prime}}^ {\rm T}, \ldots,
{{\bm v}_N^{\prime}}^ {\rm T} )^{\rm T}
\label{aeB7}
\end{equation}
with
\begin{equation}
{\bm r}_i^{\prime}={\bm r}_i - \langle{\bm r}_i \rangle ~~{\rm and}~~
{\bm v}_i^{\prime}={\bm v}_i - \langle{\bm v}_i \rangle~,
\label{aeB8}
\end{equation}
where ${\bm r}_i$ and ${\bm v}_i$ are the coordinate and velocity vectors
of the $i$th atom of the biopolymer,
and $\langle {\bm r}_i \rangle$ and $\langle {\bm v}_i \rangle$ are
the average coordinate and velocity vectors.
The operators $e^{-\iGa^{\dagger} t_0/2}$
and $e^{-\iGa t_0/2}$ are used to reduce the contributions of
faster modes in ${\bm R}$ and ${\bm R}P_{\rm eq}$, respectively, and the
trial functions become better approximations as $t_0$ becomes larger.
\par
For the trial functions Eqs.\ (\ref{aeB5}) and (\ref{aeB6}),
the variational equation is
equivalent to a generalized eigenvalue problem
\begin{equation}
{\bm f}_p^{\rm T} C(t_0+\tau)
= {\rm e}^{-\lambda_p \tau}
{\bm f}_p^{\rm T} C(t_0)
\label{aeB9-1}
\end{equation}
and
\begin{equation}
C(t_0+\tau){\hat{\bm f}_p}
= {\rm e}^{-\lambda_p \tau}
C(t_0){\hat{\bm f}_p}
,
\label{aeB9-2}
\end{equation}
where $\lambda_p$ is the relaxation rate corresponding to $X_p$ and
$\hat{X}_p$.
The matrix $C(t)$ is an equilibrium time-displaced correlation matrix defined by
\begin{equation}
C(t)=
\langle {\bm R}(t) {\bm R}^{\rm T}(0) \rangle~.
\label{aeB10}
\end{equation}
The orthonormal condition for $X_{p}$ and $\hat{X}_{p}$
is written as
\begin{equation}
{\bm f}_p^{\rm T}
C(t_0)
{\hat{\bm f}_q}
=\delta_{p,q}.
\label{aeB11}
\end{equation}
The time correlation functions of relaxation modes are now written as
\begin{equation}
	\langle X_p(t)\hat{X}_q(0) \rangle \simeq \delta_{p,q}\exp(-\lambda_p t)~.
\end{equation}
Because $X_P$ and $\hat{X}_q$ are approximate relaxation modes
determined by solving    for one value of $\tau$, the relation holds exactly for
$t=0$ and $t=\tau$, and is expected to hold approximately
for other values of $t \ge 0$.
\par
By using Eq.\ (\ref{aeA20}),
we choose
\begin{equation}
 \hat{X}_p(Q) = X_p(\epsilon Q).
 \end{equation}
Then, the detailed balance condition (\ref{db}) leads to
\begin{equation}
\hat{\bm f}_p = E {\bm f}_p
\label{fhat=Ef}
\end{equation}
with
\begin{equation}
E = {\rm diag}(
\epsilon_1,\epsilon_1,\epsilon_1,
\epsilon_2,\epsilon_2,\epsilon_2,
\ldots,
\epsilon_{2N},\epsilon_{2N},\epsilon_{2N}
).
\end{equation}
Because the detailed balance condition also leads to
\begin{equation}
	C(t)^{\rm T} = E C(t) E,
\label{CT=ECE}
\end{equation}
it can be seen that
the generalized eigenvalue problems (\ref{aeB9-1}) and (\ref{aeB9-2})
are equivalent.
\par
From the orthonormal condition, the inverse transformation
of Eqs.\ (\ref{aeB5}) and (\ref{aeB6}) is derived as
\begin{equation}
{\rm e}^{-\Gamma^\dagger t_0/2}{\bm R}
=
\sum_{p=1}^{6N-12} {\bm g}_p X_p
\label{aeB12}
\end{equation}
and
\begin{equation}
P_{\rm eq}^{-1}{\rm e}^{-\Gamma t_0/2}{\bm R}P_{\rm eq}
=
\sum_{p=1}^{6N-12} \hat{{\bm g}}_p \hat{X}_p
\label{aeB13}
\end{equation}
with
\begin{equation}
{\bm g}_p= C(t_0)\hat{{\bm f}}_p
~~{\rm and}~~
\hat{{\bm g}}_p
= C(t_0)^{\rm T}{\bm f}_p.
\label{aeB14}
\end{equation}
In Eqs.\ (\ref{aeB12}) and (\ref{aeB13}),
it is assumed that the translational and rotational motions are removed.
It follows from Eqs.\ (\ref{fhat=Ef}) and (\ref{CT=ECE}) that
\begin{equation}
\hat{{\bm g}}_p = E {\bm g}_p.
\end{equation}
\par
By using Eqs.\ (\ref{aeB12}) and (\ref{aeB13}),
the equilibrium time-displaced correlation functions of ${\bm R}$ are given as
\begin{align}
	C(t)=\langle {\bf R}(t) {\bf R}^{\rm T}(0) \rangle
	&=
	\sum_{p=1}^{6N-12} \sum_{q=1}^{6N-12}
	{\bm g}_p {\bm {\hat g}}^{T}_q \langle X_p(t-t_0)X_q(0) \rangle
	\nonumber\\
	& \simeq
	\sum_{p=1}^{6N-12}
	\tilde{{\bm g}}_{p} \tilde{{\bm {\hat g}}}^{T}_{p}
	{\rm e}^{-\lambda_p t}
\label{aeB15}
\end{align}
for $t \ge t_0$.
Here,
\begin{equation}
{\tilde{\bm g}}_{p}=e^{\lambda_p t_0/2} {\bm g}_{p} ~~{\rm and }~~
{\tilde{\bm {\hat g}}}_{p}=e^{\lambda_p t_0/2}{\bm {\hat g}}_{p}~.
\label{gtilde}
\end{equation}
Note that
\begin{equation}
{\tilde{\bm {\hat g}}}_{p} = E {\tilde{\bm g}}_{p}.
\end{equation}
Equation (\ref{aeB15}) can be regarded as Eq.\ (\ref{aeA18})
with the mode expansion
\begin{equation}
	{\bm R} \simeq  \sum_{p=1}^{6N-12} {\tilde{\bm g}}_{p}X_p
	~~{\rm and}~~
	{\bm R} \simeq  \sum_{p=1}^{6N-12} {\tilde{\bm {\hat g}}}_{p} \hat{X}_p.
\end{equation}
\par
The RMA method explained above can be summarized as follows.
The approximate relaxation modes $X_p$
and $\hat{X}_q$, which are specified by
${\bm f}_{p}$ and $\hat{\bm f}_{p}= E{\bm f}_p$
through Eqs. (\ref{aeB5}) and (\ref{aeB6}),
and the corresponding relaxation rate $\lambda_p$ are determined by
solving the eigenvalue problem
(\ref{aeB9-1}) or (\ref{aeB9-2}),
which are equivalent,
with the condition (\ref{aeB11}),
where
the equilibrium time-displaced correlation matrix
$C(t)$ is defined by (\ref{aeB10}).
In terms of the approximate relaxation modes and rates,
$C(t)$ with $t \ge t_0$ can be calculated from Eq.\ (\ref{aeB15}),
where $\tilde{\bm g}_p$ and $\tilde{\hat{\bm g}}_p = E \tilde{\bm g}_p$
are given by Eq.\ (\ref{gtilde}).
By comparing $C(t)$ calculated in the manner described above with that directly calculated by
simulations,
the accuracy of the approximate relaxation modes and rates
can be examined.
\par
In this paper,
we deal with position coordinates only, for which $\epsilon_i = 1$.
Therefore,
the following relations hold:
$X_p = \hat{X}_p$,  ${\bm f}_p = \hat{\bm f}_p$,
${\bm g}_p = \hat{\bm g}_p$,
$\tilde{{\bm g}}_p = \tilde{\hat{\bm g}}_p$,
and
$C(t)=C(t)^{\rm T}$.

\newpage

\noindent
{\bf \Large FIGURE CAPTIONS:}\\


%
\newpage
\begin{figure}
\includegraphics[width=8cm]{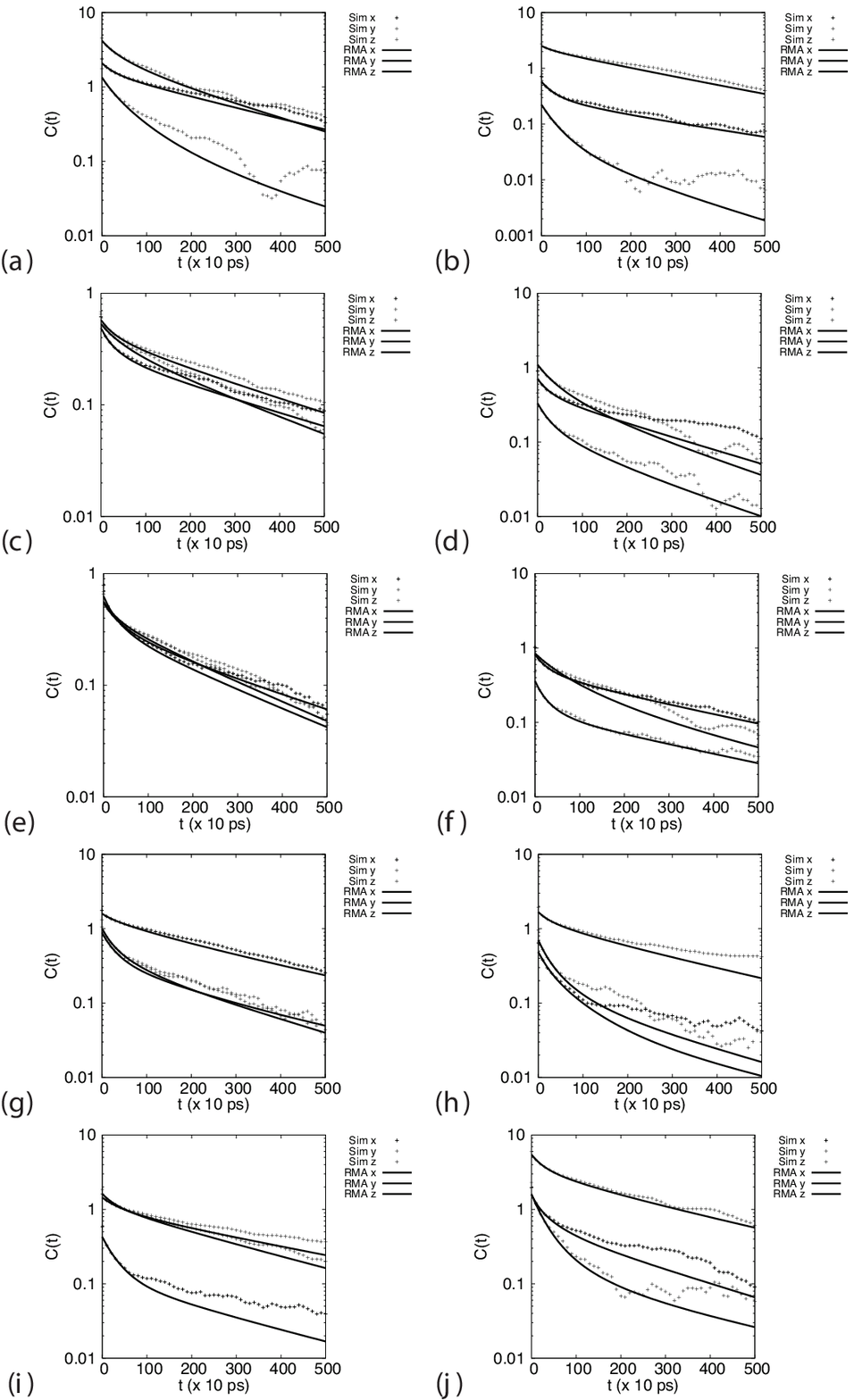}
\label{FigS1}
\yfigcaption{{\bf Fig. S1 (Supplemental Material): }
The time-displaced autocorrelation functions of x,y,z-coordinates for $i$th C$_\alpha$ atom ($i=1,\cdots,10$)  
obtained by the simulation directly (Sim) and reproduced by RMA (RMA). }
\end{figure}
\newpage
\begin{figure}
\begin{center}
\includegraphics{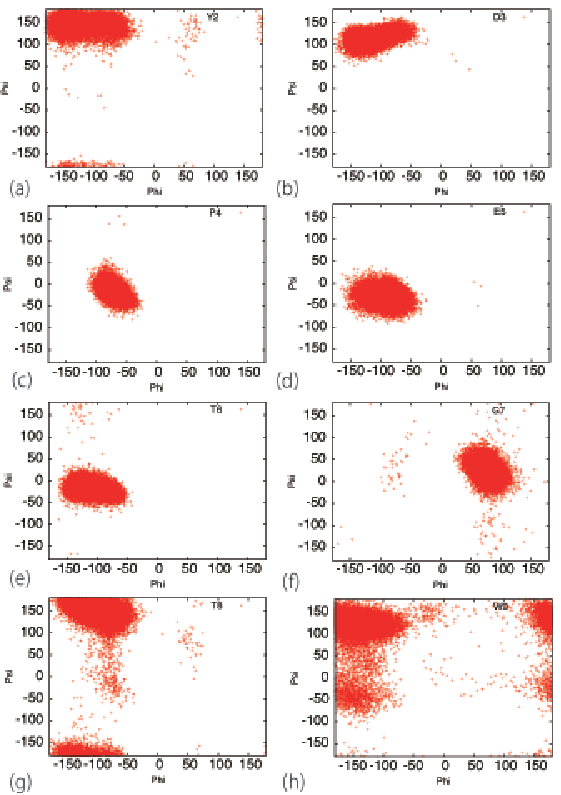}
\end{center}
\yfigcaption{ {\bf Fig. S2-1 (Supplemental Material): } 
The ramachandran plots of each residue from Tyr2 to Trp9 (from (a) to (h)) 
of the nateve state.}
\end{figure}
\newpage
\begin{figure}
\begin{center}
\includegraphics{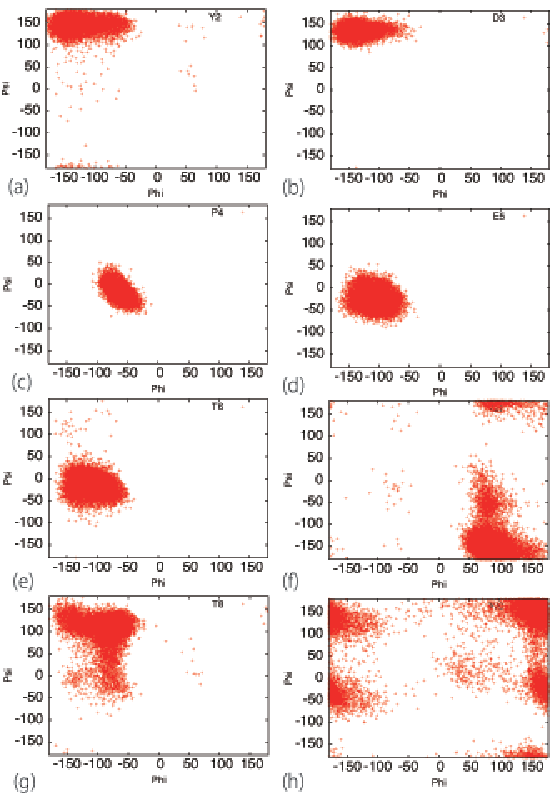}
\end{center}
\yfigcaption{ {\bf Fig. S2-2 (Supplemental Material): } 
The ramachandran plots of each residue from Tyr2 to Trp9 (from (a) to (h)) 
of the misfolded state.}
\end{figure}
\newpage
\begin{figure}
\begin{center}
\includegraphics{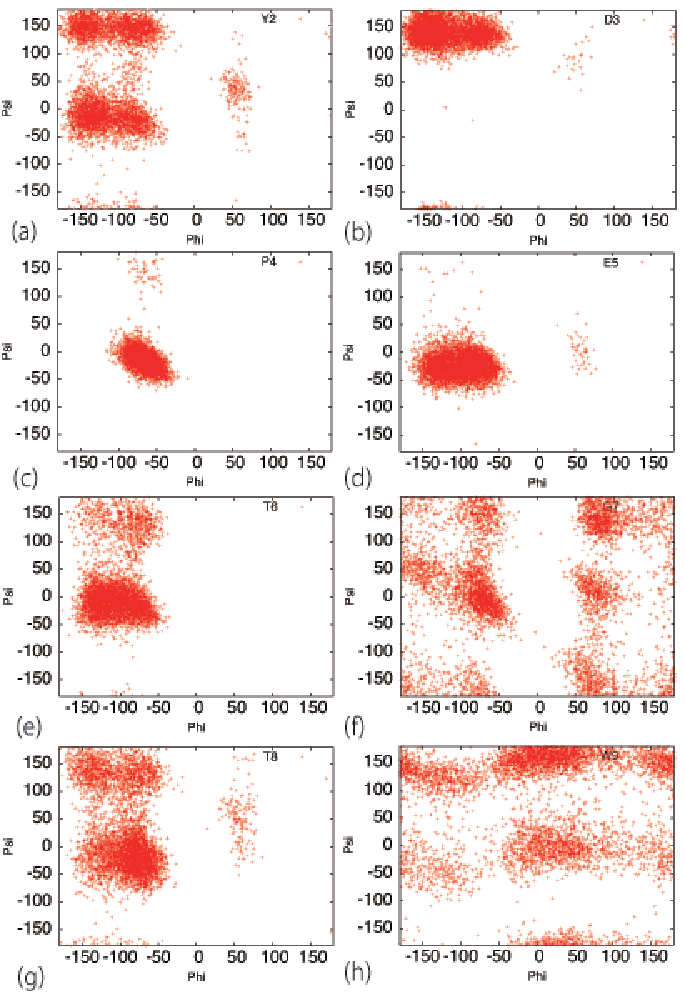}
\end{center}
\yfigcaption{ {\bf Fig. S2-3 (Supplemental Material): } 
The ramachandran plots of each residue from Tyr2 to Trp9 (from (a) to (h)) 
of the intermediate state.}
\end{figure}
\newpage
\begin{figure}
\begin{center}
\includegraphics{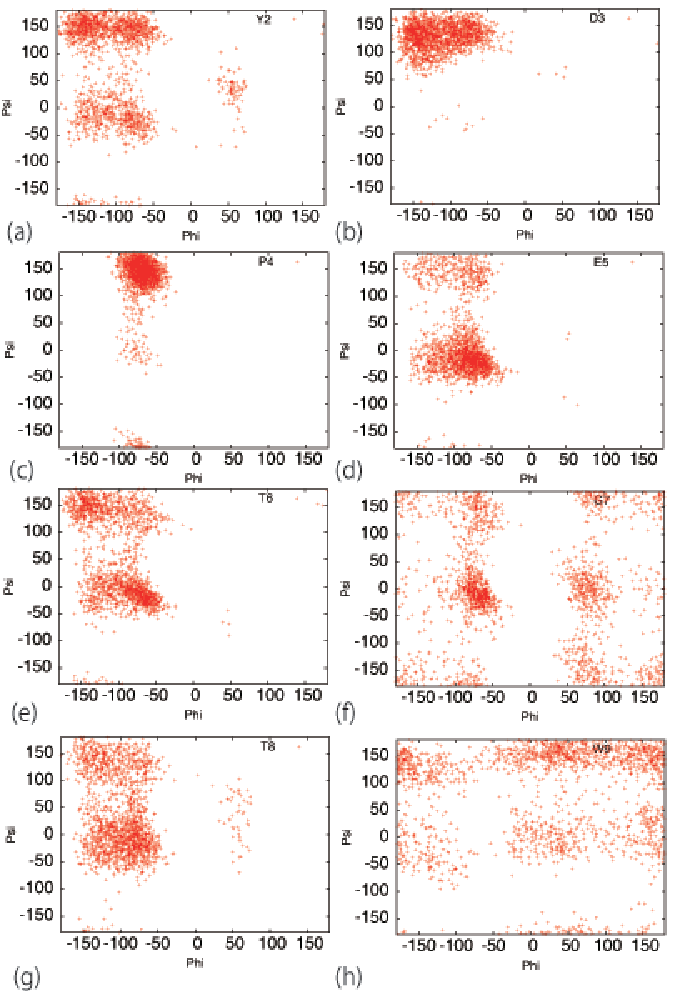}
\end{center}
\yfigcaption{ {\bf Fig. S2-4 (Supplemental Material): } 
The ramachandran plots of each residue from Tyr2 to Trp9 (from (a) to (h)) 
of the unfolded state.}
\end{figure}
\newpage
\begin{figure}
\includegraphics[width=8cm]{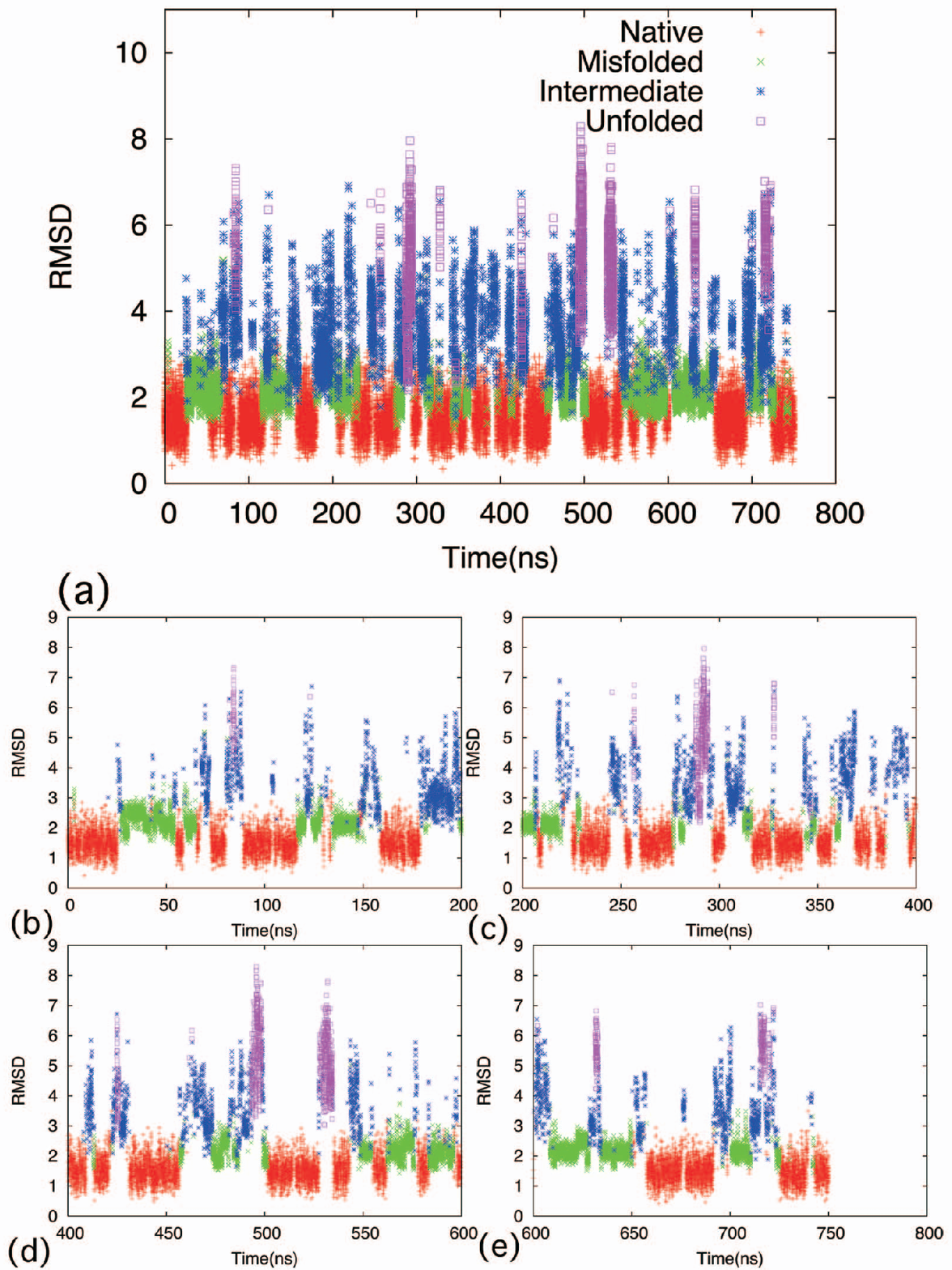}
\yfigcaption{ {\bf Fig. S3 (Supplemental Material): } 
The time series of the values of RMSD from the native structure for the native, misfolded, intermediate, and unfolded states listed in Table II. The time series of all steps (a), from 200 ns to 400 ns (b), 
from 400 ns to 600 ns (c), and from 600 ns to 800 ns (d) are shown. }
\end{figure}


\noindent
\begin{thebibliography}{(00)}
\bibitem{Shaw} K. Lindorff-Larsen, S. Piana, R. O. Dror, and D. E. Shaw,
{\it Science} {\bf 334}, 517 (2011): R. O. Dror, R. M. Dirks, J. P. Grossman, H. Xu, and D. E. Shaw, {\it Annu. Rev. Biophys.} {\bf 41}, 429 (2012).
\bibitem{Pande}T. J. Lane, D. Shukla, K. A. Beauchamp, and V. S. Pande,
{\it Curr. Opin. Struct. Biol.} {\bf 23}, 58 (2013).
\bibitem{nphys} P. L. Freddoline, C. B. Harrison, Y. Liu, and K. Schulten, {\it Nature Physics} {\bf 6}, 751 (2010).

\bibitem{KHG} A. Kitao, F. Hirata, and N. Go, {\it Chem. Phys.}, {\bf 158}, 447--472 (1991).
\bibitem{IK} T. Ichiye and M. Karplus, {\it Proteins}, {\bf 11}, 205--217 (1991).
\bibitem{AA} R. Abagyan and P. Argos, {\it J. Mol. Biol.}, {\bf 225}, 519--532 (1992).
\bibitem{G} A.E. Garcia, {\it Phys. Rev. Lett.}, {\bf 68}, 2696--2699 (1992).
\bibitem{HKHG} S. Hayward, A. Kitao, F. Hirata, and N. Go,
{\it J. Mol. Biol.}, {\bf 234}, 1207--1217 (1993).
\bibitem{ALB} A. Amadei, A.B.M. Linssen, and H.J.C. Berendsen,
{\it Proteins} {\bf 17}, 412--425 (1993).
\bibitem{qh} R. M. Levy, A. R. Srinivasan, W. K. Olson, and J. A. McCammon,
{\it Biopolymers}, {\bf 23}, 1099-1112 (1984).
%

\bibitem{TM} H. Takano and S. Miyashita, {\it J. Phys. Soc. Jpn.}, {\bf 64}, 3688--3698 (1995).
\bibitem{KHT} S. Koseki, H. Hirao, and H. Takano, {\it J. Phys. Soc. Jpn.}, {\bf 66}, 1631--1637 (1997).
\bibitem{HKT} H. Hirao, S. Koseki, and H. Takano, {\it J. Phys. Soc. Jpn.}, {\bf 66}, 3399--3405 (1997).
\bibitem{HT1}
K.\ Hagita and H.\ Takano,
{\it J.\ Phys.\ Soc.\ Jpn.},
{\bf 68}, 401--407 (1999).
\bibitem{HaKT}
K.\ Hagita, S.\ Koseki and H.\ Takano,
{\it J.\ Phys.\ Soc.\ Jpn.},
{\bf 68}, 2144--2145 (1999).
\bibitem{HIT}
K.\ Hagita, D.\ Ishizuka and H.\ Takano,
{\it J.\ Phys.\ Soc.\ Jpn.},
{\bf 70}, 2897--2902 (2001).
\bibitem{HT2}
K.\ Hagita and H.\ Takano,
{\it J.\ Phys.\ Soc.\ Jpn.},
{\bf 71}, 673--676 (2002).
\bibitem{HT3}
K.\ Hagita and H.\ Takano,
{\it J.\ Phys.\ Soc.\ Jpn.},
{\bf 72}, 1824--1827 (2003)
and references therein.
\bibitem{ST}
S.\ Saka and H.\ Takano,
{\it J.\ Phys.\ Soc.\ Jpn.},
{\bf 77}, 034001 (2008).
\bibitem{IT}
N.\ Iwaoka and H.\ Takano,
{\it J.\ Phys.\ Soc.\ Jpn.},
{\bf 82}, 064801 (2013).
\bibitem{IT2}
N.\ Iwaoka and H.\ Takano,
{\it J.\ Phys.\ Soc.\ Jpn.},
{\bf 83}, 123801 (2014).
\bibitem{IHT}
N.\ Iwaoka, K.\ Hagira, and H.\ Takano,
{\it J.\ Phys.\ Soc.\ Jpn.},
{\bf 84}, 044801 (2015).
\bibitem{dG}
P.\ G.\ de Gennes,
{\it Scaling Concepts in Polymer Physics}
(Cornell University Press, Ithaca, 1984).
\bibitem{DE}
M.\ Doi and S.\ F.\ Edwards,
{\it The Theory of Polymer Dynamics}
(Oxford University Press, Oxford, 1986).
\bibitem{MIT}
A. Mitsutake, H. Iijima, and H. Takano,
{\it J. Chem. Phys. },
{\bf 135}, 164102 (2011).
\bibitem{NMT} T.\ Nagai, A.\ Mitsutake, and H.\ Takano,
{\it J. Phys. Soc. Jpn.}, {\bf 82}, 023803 (2013);
T.\ Nagai, A.\ Mitsutake, and H.\ Takano,
Seibutsu Butsuri (Biophysics), {\bf 49}, Supplement S75,
(Abstracts for the 47st annual meeting, The Biophysical Society of Japan) (2009).
\bibitem{R}
H.\ Risken,
{\it The Fokker-Planck Equation: Methods of Solution and Applications},
2nd Ed.\ (Springer-Verlag, Berlin, 1989).

\bibitem{SPS}
W.C. Swope, J.W. Pitera, and F. Suits, {\it J. Phys. Chem.B}, {\bf 108}, 6571 (2004).
\bibitem{SP} N. Singhal, C. Snow, and V.S. Pande, {\it J. Chem. Phys.},{\bf 121}, 415 (2004).
\bibitem{CSPD}
J.D. Chodera, W.C. Swope, J.W. Pitera, and K.A. Dill,
{\it Multiscale Model. Simul.} {\bf 5}, 1214 (2006).
\bibitem{CDSPSW}
J.D. Chodera, K.A. Dill, N. Singhal, V.S. Vande, W.C. Swope, and J.W. Pitera,
{\it J. Chem. Phys}. {\bf 126}, 155101 (2007).
\bibitem{NHSS}
F. No\'{e}, U. Horenko, C. Sch\"{u}tte, and J.C. Smith, {\it J. Chem. Phys.} {\bf 126}, 155102 (2007).
\bibitem{BH}
N. Buchete and G. Hummer, {\it J. Phys. Chem.B} {\bf 112}, 6057 (2008).

\bibitem{MSM}
F. No\'{e} and S. Fischer, {\it Curr. Opin. Struct. Biol.}, {\bf 18}, 154 (2008).
\bibitem{NoeRev}
J. D. Chodera and F. No\'{e}, {\it Curr. Opin. Struct. Biol.}, {\bf 25}, 135 (2014).
\bibitem{PandeRev}
C.R. Schwantes, R.T. McGibbon, and V.S. Pande, {\it J. Chem. Phys.}, {\bf 141}, 090901 (2014).

\bibitem{Fuchigami}
Y. Naritomi and S. Fuchigami, {\it J. Chem. Phys.}, {\bf 134}, 065101 (2011).
\bibitem{MSM2}
G. Pe'rez-Hernandez, F. Paul, T. Giorgino, G. D. Fabritiis, and F. No\'{e}, {\it J. Chem. Phys.}, {\bf 139}, 015102 (2013).
\bibitem{MSMPANDE}
C. R. Schwantes and V. S. Pande, {\it J. Chem. Theo. Comput.}, {\bf 9}, 2000 (2013).

\bibitem{chig} S. Honda, K. Yamasaki, Y. Sawada, H. Morii, {\it Structure} {\bf 12}, 1507 (2004).
\bibitem{terada} D. Satoh, K. Shimizu, S. Nakamura, and T. Terada, {\it FEBS Letters}, {\bf 580}, 3422 (2006).
\bibitem{taiji} A. Suenaga, T. Narumi, N. Futatsugi, R. Yanai, Y. Ohno, N. Okimoto, and M. Taiji,
{\it Chem. Asian J.} {\bf 2} 591 (2007).
\bibitem{haradakitao}
R. Harada and A. Kitao, {\it J. Phys. Chem. B}, {\bf 115}, 8806 (2011).
\bibitem{Best}
P. Kuhrova, A. D. Simone, M. Otyepka, and R. B. Best,
{\it Biophys. J.}, {\bf 102}, 1897 (2012).
\bibitem{Oku}
H. Okumura, {\it Proteins }, {\bf 80}, 2397 (2012).

\bibitem{AMBER8}
D.\ A.\ Case, T.\ A.\ Darden, T.\ E.\ Cheatham, III, C.\ L.\ Simmerling, J.\ Wang,
R.\ E.\ Duke, R.\ Luo, K.\ M. \ Merz, B.\ Wang, D.\ A.\ Peariman, M.\ Crowley,
S.\ Brozell, V.\ Tsui, H.\ Gohlke, J.\ Mongan, V.\ Hornak, G.\ Cui,
P.\ Beroza, C.\ Schafmeister, J.\ W.\ Caldwell, W.\ S.\ Ross,
and P.\ A.\ Kollman, AMBER8, University of California, San Francisco (2004).

\bibitem{ROT}
C. Eckart, {\it Phys. Rev.}, {\bf 47}, 552 (1935).
\bibitem{ROT2}
A. D. McLachlan,
{\it J. Mol. Biol.}, {\bf 128}, 49 (1979).

\bibitem{supple} See supplemental material at [URL will be inserted by AIP]
for additional results of Figs. S1, S2, and S3.

\bibitem{vmd}
W.\ Humphrey, A.\ Dalke, and K.\ Schulten,
{\it J. Mol. Graph.}, {\bf 14}, 33 (1996).
\end{thebibliography}
\end{document}